\documentclass[a4paper,12pt]{article}
\usepackage{amsfonts,amsthm,amsmath,amssymb,graphicx,hyperref,youngtab}

\def\Tr{{\rm Tr\, }}

\def\b{\beta}

\def\G{\Gamma}

\def\m{\mu}

\def\n{\nu}

\def\s{\sigma}

\newcommand{\be}{\begin{equation}}
\newcommand{\bea}{\begin{eqnarray}}
\newcommand{\ee}{\end{equation}}
\newcommand{\eea}{\end{eqnarray}}

\newcommand{\nn}{\nonumber}

\def\s{ \sigma}

\begin{document}

\makeatletter
\@addtoreset{equation}{section}
\makeatother
\renewcommand{\theequation}{\thesection.\arabic{equation}}

\rightline{WITS-CTP-114}
\vspace{1.8truecm}

\vspace{15pt}


{\LARGE{  
\centerline{   \bf Operators, Correlators and Free Fermions} 
\centerline   {\bf  for $SO(N)$ and $Sp(N)$} 
}}  

\vskip.5cm 

\thispagestyle{empty} \centerline{
    {\large \bf Pawel Caputa$^a$\footnote{{\tt pawel.caputa@wits.ac.za}}, Robert de Mello Koch$^{a,b}$\footnote{{\tt robert@neo.phys.wits.ac.za}} }
   {\large \bf and Pablo Diaz$^a$\footnote{{\tt Pablo.DiazBenito@wits.ac.za}} }}

\vspace{.4cm}
\centerline{{\it $^{a}$National Institute for Theoretical Physics,}}
\centerline{{\it Department of Physics and Centre for Theoretical Physics,}}
\centerline{{\it University of Witwatersrand, Wits, 2050, } }
\centerline{{\it South Africa } }
\vspace{.4cm}
\centerline{{\it $^{b}$Institute of Advanced Study,}}
\centerline{{\it Durham University}}
\centerline{{\it Durham DH1 3RL, UK}}
\vspace{.4cm}

\thispagestyle{empty}

\centerline{\bf ABSTRACT}

\noindent
Using the recently constructed basis for local operators in free $SO(N)$ gauge theory 
we derive an exact formula for the correlation functions of multi trace operators.
This formula is used to obtain a simpler form and a simple product rule 
for the operators in the $SO(N)$ basis.
The coefficients of the product rule are the Littlewood-Richardson numbers
which determine the corresponding product rule in free $U(N)$ gauge theory. 
$SO(N)$ gauge theory is dual to a non-oriented string theory on the 
AdS$_5\times {\mathcal R}$P$^5$ geometry. To explore the physics 
of this string theory we consider the limit of the gauge theory that,
for the $U(N)$ gauge theory, is dual to the pp-wave
limit of $AdS_5 \times S^5$. Non-planar unoriented ribbon diagrams do not survive this 
limit. We give arguments that the number of operators in our basis matches counting
using the exact free field partition function of free $SO(N)$ gauge theory. 
We connect the basis we have constructed to free fermions,
which has a natural interpretation in terms of a class of ${1\over 2}$-BPS bubbling 
geometries, which arise as orientifolds of type IIB string theory.
Finally, we obtain a complete generalization of these results to $Sp(N)$ gauge theory
by proving that the finite $N$ physics of $SO(N)$ and $Sp(N)$ gauge theory are related 
by exchanging symmetrizations and antisymmetrizations and replacing $N$ by $-N$.

\vskip.4cm 

\setcounter{page}{0}
\setcounter{tocdepth}{2}

\newpage

\tableofcontents

\setcounter{footnote}{0}

\linespread{1.1}
\parskip 4pt

{}~
{}~

\section{Introduction}

In a previous paper\cite{firstpaper} we have initiated the study of local operators in $SO(N)$ gauge theory which have a bare dimension that can
depend parametrically on $N$.
For these operators, one needs to sum more than just the planar diagrams to capture the large $N$ limit.
We dealt with this problem by employing group representation theory to define local operators which generalize the Schur polynomials of the 
theory with gauge group $U(N)$. 
We found that the free field two point function is diagonalized by our operators.
In this article we will extend our understanding in a number of important ways.

A basic result of \cite{firstpaper} is the basis of local operators, given by
\bea
  O_R (Z) ={1\over (2n)!}\sum_{\sigma\in S_{2n}}\chi_R(\sigma )
           \sigma^{ i_1 i_2 i_2 i_1\cdots i_{n-1 }i_n i_n i_{n-1}}_{j_1 j_2 \cdots j_{2n-1} j_{2n}}
            Z^{j_1 j_2}Z^{j_3 j_4}\cdots Z^{j_{2n-1} j_{2n}}
\label{opone}
\eea
For $SO(N)$ with $N$ even, we also need to include\footnote{Recall that there are two invariant tensors for $SO(N)$: the
Kronecker delta $\delta_{ij}$ and the $\epsilon_{i_1 i_2\cdots i_N}$. We can use either of these tensors when contracting indices to obtain
gauge invariant operators. For $SO(N)$ with $N$ odd $\epsilon_{i_1 i_2\cdots i_N}$ has an odd number of indices, so that we can't build a
gauge invariant operator that uses only a single $\epsilon_{i_1 i_2\cdots i_N}$.}
\bea
Q_R (Z)&=&{\epsilon_{i_1 i_2\cdots i_N} \over (N+4)!}\sum_{\sigma\in S_{N+2p}}\chi_R(\sigma)
\sigma^{i_1 i_2\cdots i_{N-1}i_N i_{N+1} i_{N+2} i_{N+2} i_{N+1}\cdots i_{N+p-1} i_{N+p} i_{N+p} i_{N+p-1}}_{j_1 j_2\cdots j_{N+2p-1} j_{N+2p}}\cr
       &&  \times Z^{j_1 j_2}Z^{j_3 j_4}\cdots Z^{j_{N+2p-3} j_{N+2p-2}}Z^{j_{N+2p-1} j_{N+2p}}
\label{optwoq}
\eea
We will focus our discussion on the $O_R(Z)$ which we understand better than the $Q_R (Z)$.
The operator label $R$ for $O_R$ is a Young diagram with $2n$ boxes, that is, an irreducible representation of the symmetric group $S_{2n}$.
To obtain a non-zero operator $O_R(Z)$, $n$ must be even. Thus, $2n$ is divisible by $4$. 
In fact, the only representations $R$ which lead to a non-zero operator are built from the ``basic block'' ${\tiny \yng(2,2)}$
as we now explain.
Choose a partition of $n/2$, or equivalently a Young diagram $r$ with $n/2$ boxes.
The set of valid representations $R$ are obtained by replacing each box in $r$ by the ``basic block'' ${\tiny \yng(2,2)}$.
To reflect the relation between $r$ and $R$, we use the notation $r=R/4$.
Thus, the number of gauge invariant operators of the type $O_R(Z)$
built using $n$ fields is equal to the number of partitions of ${n\over 2}$.
As an example, for $n=4$ the allowed labels are
\bea
  R_1/4 = \yng(2),\qquad R_2/4 = \yng(1,1)
\eea
\bea
  R_1 = \yng(4,4),\qquad R_2 = \yng(2,2,2,2)
\eea
Given the form of the Young diagrams $R$ we consider, we will list the row lengths of $R$ as $(r_1,r_1,r_2,r_2,\cdots,r_{N\over 2},r_{N\over 2})$.
In section 2 we start by using the results of \cite{firstpaper} to derive an exact formula for the free field theory correlation 
functions of multi trace operators.
The result is given in equation (\ref{completeresult}).
Using this formula we explain in section 3, how to obtain a simpler form for the $O_R(Z)$ of our $SO(N)$ basis. 
The resulting operators, with a more convenient normalization, are
\bea
  \chi_S^0 (Z)= {1\over \left({n\over 2}\right)!}
      \sum_{\nu\in S_{n\over 2}}2^{-l(\nu)}\chi_{S/4}(\nu){\rm Tr}_{V^{\otimes\, n}}(\nu (Z^2)^{\otimes {n\over 2}})
\eea
With the new normalization, the two point function of these operators is
\bea
   \langle \chi_R^0 (Z) \chi_S^0 (\bar{Z})\rangle =\delta_{RS}\prod_{i\in {\rm odd\, boxes\, in\,}S} c_i
\eea
A few comments are in order.
Each box in a Young diagram can be assigned a factor, denoted $c_i$ in the above formula. 
A box appearing in column $a$ and row $b$ has factor $N+a-b$.
The right hand side is equal to the product of the factors of the boxes in every second row as shown below
\bea
  \young(\,\,\,\,,****,\,\,,**)
\eea
We called these the odd boxes in \cite{firstpaper} because they referred to boxes labeled with an odd integer in a Young-Yamanouchi
labeling of the states in the $S_{2n}$ irreducible representation.
With the new simplified form of the $SO(N)$ operators, we are able, in section 3, to give a product rule for our operators.
The product rule is ($S/4\vdash {n_1\over 2}$, $R/4\vdash {n_2\over 2}$)
\bea
     \chi_S^0(Z)\chi_R^0(Z) = \sum_{T/4\vdash {n_1+n_2\over 2}}\, g_{R/4\,S/4\,T/4} \chi_T^0(Z)
\eea
where $g_{R/4\,S/4\,T/4}$ is the Littlewood-Richardson coefficient. 

These results constitute a rather complete understanding of the local operators in $SO(N)$ gauge theory, 
comparable to what has been achieved for the $U(N)$ theory.
This program was initiated in the context of $U(N)$ gauge theory, by Corley, Jevicki and Ramgoolam in \cite{cjr}.
In particular, \cite{cjr} showed that the half-BPS operators constructed using a single complex matrix can be described using Schur polynomials
and they demonstrated that the Schur polynomials diagonalize the free field two point function.
The study of the finite $N$ physics of $U(N)$ gauge theories is by now well developed.
There are a number of bases of local operators that diagonalize the free field two point 
function\cite{dssi,Kimura:2007wy,BHR1,BHR2,Bhattacharyya:2008rb,Kimura:2009jf,Kimura:2012hp,jurgis}
and we know how to diagonalize the one-loop dilatation operator\cite{Koch:2010gp,DeComarmond:2010ie} 
for certain operators dual to giant graviton branes\cite{Carlson:2011hy,gs,Koch:2011hb,DCI,Berenstein:2013md}. 
This diagonalization has provided new integrable sectors, with the spectrum of the dilatation operator reducing to that of 
decoupled harmonic oscillators which describe the excitations of the system\cite{gs,Koch:2011hb,DCI}. 
Integrability in the planar limit was discovered in \cite{mz,bks} and is reviewed in \cite{intreview}.
For a study of the $SU(N)$ theory see \cite{gwyn}.

Given the results we have developed, we are now in a position to probe the finite $N$ physics of $SO(N)$ gauge theory.
Recall that according to the AdS/CFT duality\cite{malda,Gubser:1998bc,Witten:1998qj}, finite $N$ physics of the gauge 
theory\cite{Balasubramanian:2001nh} corresponds to non-perturbative (in the string coupling) physics of 
objects such as giant graviton branes\cite{mst,myers,hash} and the stringy exclusion principle\cite{Maldacena:1998bw}.
${\cal N}=4$ super Yang-Mills with $SO(N)$ or $Sp(N)$ gauge group is dual to the AdS$_5\times {\mathcal R}$P$^5$ geometry\cite{Witten:1998xy}.
In this case one expects a non-oriented string theory so that the study of non-perturbative stringy physics,
which is captured by the finite $N$ physics of the gauge theory, is likely to provide new insights extending
what can be learned from the AdS$_5\times$S$^5$ example which involves oriented string.
For studies in this direction see \cite{Aharony:2002nd}. 
Computations in the gauge theory, must sum both the planar and the non-planar diagrams.
At the non-planar level there are genuine differences between the $U(N)$ and the $SO(N)$ or $Sp(N)$ gauge theories.
Recall that matrix model Feynman diagrams in double line notation represent discrete triangulations of Riemann surfaces\cite{'tHooft:1973jz}. 
For Hermitian matrices we deal with oriented triangulations whereas for 
symplectic or anti-symmetric matrices unoriented triangulations\cite{Cicuta:1982fu}. 
In general, a Feynman diagram is weighted by
\be
\lambda^{2g-2+b+c}N^{-c-2g+2},
\ee
where $N$ is the number of colors, $\lambda=g^{2}_{YM}N$ is the 't Hooft coupling, $g$ is the number of handles, $b$ the number of 
boundaries and $c$ the number of cross-caps on the surface.
Thus, for the $SO(N)$ or $Sp(N)$ gauge theories, the leading non-planar corrections come from
ribbon graphs that triangulate non-orientable Feynman diagrams with a single cross-cap.
The large $N$ limit of correlation functions of operators with a bare dimension that depends parametrically on $N$
are sensitive to this non-planar structure of the theory. 
With the goal of probing this structure, in section 4 we use our technology to compute free field theory correlation
functions of multi trace operators, to all orders in $1/N$.
Given these correlators, we can consider the double scaling limit defined by \cite{Berenstein:2002jq,Kristjansen:2002bb}
\begin{equation}
N\to\infty~~\text{and}~~J\to\infty~~\text{with}~~\frac{J^2}{N}~~\text{fixed},~~g_{YM}~~\text{fixed}\label{DS}
\end{equation}
where $J$ is the number of fields in the gauge theory operator. In this limit some non-planar diagrams (string interactions) 
survive, giving a non-trivial normalization of the two and three-point correlators of single trace operators.
This limit is particularly interesting because in the dual gravity picture it corresponds to taking a pp-wave limit of $AdS_5 \times S^5$,
a background in which the superstring theory can be quantized. 
We find that non-planar unoriented diagrams in $SO(N)$ gauge theory do not survive this limit.

We have argued that $O_R$ (and for $N$ even, the $Q_R$) give a basis. 
A weak point in our argument is that we have not demonstrated that these operators are a complete set. 
This issue is considered in section 5. 
We focus on $N$ even, which is the more involved case, as a consequence of the fact that we may use $\epsilon_{i_1 i_2\cdots i_N}$ 
when constructing gauge invariant operators. 
By counting the number of operators we have constructed, we are able to reproduce the exact free field partition function
of the $SO(N)$ gauge theory\cite{Aharony:2003sx}
for $N=4,6$ and, using our results, we give a conjecture for the free field partition function
at any even $N$, in (\ref{exactpart}) and at any odd $N$ in (\ref{exactoddnum}).

The basis that we have constructed allows us to study the dynamics of the gauge invariant observables of a single matrix model.
Of course, this problem can be reduced to eigenvalue dynamics which is itself equivalent to the dynamics of free fermions in an
external potential\cite{BIPZ}.
The Schur polynomial basis for the $U(N)$ gauge theory has a very direct link to free fermion dynamics.
It is natural to ask if there is a similar connection between free fermions and the basis we have constructed.
We develop this link in some detail in section 6 and show that there is indeed a natural connection to free fermions.
Our operators can be mapped to states of fermions moving in a harmonic oscillator potential, with definite parity and maximum angular momentum
for a given energy.

Although we will not do so in this article, we have developed enough technology that
it would be natural to initiate a systematic study of the dilatation operator
in non-planar large $N$ limits of the $SO(N)$ and $Sp(N)$ gauge theories. 
A detailed study of the planar spectral problem of ${\cal N}=4$ super Yang-Mills with gauge 
groups $SO(N)$ and $Sp(N)$ has been carried out in \cite{Caputa:2010ep}.
The essential difference between the theories with gauge groups $U(N)$
or $SO(N)$ is that in the $SO(N)$ case certain states are projected out.
Thus, the planar spectral problem of the $SO(N)$ theory can again be mapped to an
integrable spin chain\cite{Caputa:2010ep}. 
It is interesting to ask if the new integrable sectors discovered in \cite{gs,Koch:2011hb,DCI} are also present in
large $N$ but non-planar limits of $SO(N)$ and $Sp(N)$ gauge theory.

It is well known that there is a close relationship between $SO(N)$ group theory and $Sp(N)$ group theory. 
These relations imply that the dimension of a given irreducible representation of $Sp(N)$ is equal
to that representation of $SO(N)$ with symmetrizations exchanged with antisymmetrizations (i.e. transpose the Young tableau) 
and $N$ replaced by $-N$\cite{King}. 
The QCD loop equations for $SO(N)$ gauge theory and $Sp(N)$ gauge theory in $3+1$ dimensions enjoy the same connection\cite{Mkrtchian:1981bb}.  
This same relation has been observed in two dimensional Yang-Mills theory\cite{Ramgoolam:1993hh}.
Motivated by this background, we proceed to study the finite $N$ physics of $Sp(N)$ gauge theory.
We are able to argue that precisely the same connection relates the finite $N$ physics for the orthogonal and symplectic gauge theories.
In this way, in section 7 we obtain a rather complete description of the finite $N$ physics of the $Sp(N)$ gauge theory.

In section 8 we outline some open problems that we find interesting.

\section{Multi trace Correlators}

Our goal in this section is to give an exact formula for the free field theory correlation 
functions of multi trace operators in $SO(N)$ gauge theory.
In our discussion below $n$, ${n\over 2}$ and $2n$ will enter at various points. 
The reader is encouraged to keep in mind that our operators are built using $n$ fields.

\begin{figure}[ht]%
\begin{center}
\includegraphics[width=15cm]{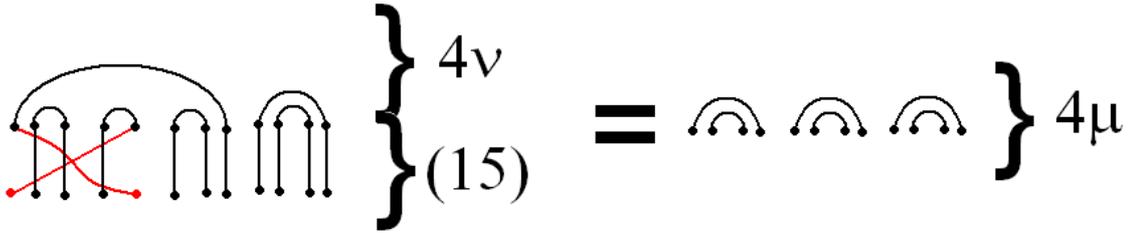}%
\caption{Relating $C^{4\nu}_I$ and $C^{4\mu}_I$ with a permutation.}%
\label{relatefig}%
\end{center}
\end{figure}

We start by choosing any partition $\nu\vdash {n\over 2}$. 
The partition is then translated into the cycle structure of a permutation.
For example, if ${n\over 2}=4$ there are 5 possible choices for $\nu$, namely $(4)$, $(2)^2$, $(2)\,(1)^2$, $(3)\, (1)$, or $(1)^4$.
We use these partitions $\nu$ (see \cite{firstpaper}) to construct $\sigma_{4\nu}\in S_{2n}$ with cycle structure given by 
multiplying each of the parts of $\nu$ by 4. 
For the above list of partitions $\nu$ the cycle structures of the corresponding permutations 
$\sigma_{4\nu}$ are (16), $(8)^2$, $(8)(4)^2$, $(12)(4)$ and $(4)^4$. 
We will now explain, by providing a few examples, how we associate to each cycle structure a canonical permutation.
The cycle structure $(12)\, (4)$ is associated to the cycle 
\bea
  \sigma_{4\nu} = (1,2,3,4,5,6,7,8,9,10,11,12)(13,14,15,16)
\eea
while $(8)\,(4)^2$ is associated to
\bea
  \sigma_{4\nu} = (1,2,3,4,5,6,7,8)(9,10,11,12)(13,14,15,16)
\eea
So, the rule for obtaining the canonical permutation is to populate the largest cycles first, starting from 1 and counting up, until the
permutation is completely determined.
Notice that the canonical permutations are composed of cycles with cycle lengths that are a multiple of 4 and further, they always take 
an even number to an odd number.
These canonical permutations can be used to define a ``contractor'' as follows
\bea
  C^{\sigma_{4\nu}}_J=C^{\sigma_{4\nu}}_{j_1 j_2\cdots j_{2n}}=\prod_{p=1}^n \delta_{j_{2p}j_{\sigma(2p)}}\cr
  C_{\sigma_{4\nu}}^J=C_{\sigma_{4\nu}}^{j_1 j_2\cdots j_{2n}}=\prod_{p=1}^n \delta^{j_{2p}j_{\sigma(2p)}}
\eea
To see that all indices appear on the right hand side and no index appears more than once, it is useful to remember that $\sigma_{4\nu}$
always takes an even number to an odd number.
Using the contractors, we can define the operators
\bea
  O^{\sigma_{4\nu}}_{R}(Z)={1\over (2n)!}\sum_{\b\in S_{2n}}\chi_R(\b)  C^{\sigma_{4\nu}}_{j_1 j_2 \cdots j_{2n}}
  \b^{j_1 j_2\cdots j_{2n}}_{i_1 i_2\cdots i_{2n-1} i_{2n}}Z^{i_1 i_2}\cdots Z^{i_{2n-1}i_{2n}}
  \label{newoperator}
\eea
These operators are particularly convenient for the question of correlation functions of multi trace operators.
Indeed, using the identity
\bea
  \delta (\sigma) ={1 \over (2n)!}\sum_{\s\in S_{2n}}d_R\chi_R (\s)
\eea
we easily find
\bea
  (Z^2)^{i_1}_{i_{\nu (1)}}(Z^2)^{i_2}_{i_{\nu (2)}}\cdots (Z^2)^{i_{n\over 2}}_{i_{\nu ({n\over 2})}} = \sum_{R} d_R O^{\sigma_{4\nu}}_{R}
\eea
Thus, any multi trace operator can easily be written as a linear combination of the $O^{\sigma_{4\nu}}_{R}$. 
Introduce the notation
\bea
  {\rm Tr}_{V^{\otimes\, n}}(\nu (Z^2)^{\otimes {n\over 2}})\equiv
  (Z^2)^{i_1}_{i_{\nu (1)}}(Z^2)^{i_2}_{i_{\nu (2)}}\cdots (Z^2)^{i_{n\over 2}}_{i_{\nu ({n\over 2})}}
\eea
Every multi trace operator can be written in this way for a suitable $\nu$.
Thus, to evaluate an arbitrary two point function of multi trace operators, all we need to do is to compute the correlator
$\langle  O^{\sigma_{4\nu}}_{R} \bar{O}^{\s_{4\mu}}_{S} \rangle$. Using the results in \cite{firstpaper}, it is straight
forward to see that
\bea
   \langle O^{\sigma_{4\nu}}_R(Z)\bar{O}^{\sigma_{4\mu}}_S(Z)\rangle
  ={\delta_{RS}\, n!2^n \over (2n)!\, d_R} \sum_{\psi\in S_{2n}} {\rm Tr}(P_{[A]}\Gamma^R(\psi))
C^{4\nu}_I(\psi)^I_JC^J_{4\mu}
\eea
We can define a permutation $\sigma_{\nu\mu}$ by the condition
\bea
  C^{4\nu}_I (\sigma_{\nu\mu})^I_J = C^{4\mu}_J
\eea
By considering an example at this point, it will be clear that $\sigma_{\nu\mu}$ defines a unique element of the double coset 
$H_{4\nu}\setminus S_{2n}/H_{4\mu}$, where $H_{4\mu}$ ($H_{4\nu}$) is a stabilizer of $\sigma_{4\mu}$ ($\sigma_{4\nu}$) respectively. 
Indeed, for
\bea
 \sigma_{4\nu}=(1,2,3,4,5,6,7,8)(9,10,11,12)
\eea
and
\bea
  \sigma_{4\mu}=(1,2,3,4)(5,6,7,8)(9,10,11,12)
\eea
we have
\bea
  \sigma_{\nu\mu}=(1,5)
\eea
See Figure \ref{relatefig} for an illustration of this example.

With the use of $\sigma_{\nu\mu}$ we can now write
\bea
   \langle O^{\sigma_{4\nu}}_R(Z)\bar{O}^{\sigma_{4\mu}}_S(Z)\rangle
  &=&{\delta_{RS}\, n!2^n \over (2n)!\, d_R} \sum_{\psi\in S_{2n}} {\rm Tr}(P_{[A]}\Gamma^R(\psi))
C^{4\mu}_I(\sigma_{\mu\nu}^{-1}\psi)^I_JC^J_{4\mu}\cr
  &=&{\delta_{RS}\, n!2^n \over (2n)!\, d_R} \sum_{\psi\in S_{2n}} {\rm Tr}(P_{[A]}\Gamma^R(\sigma_{\mu\nu} \psi))
C^{4\mu}_I(\psi)^I_JC^J_{4\mu}
\eea
Since we are contracting row and column labels of $\psi$ in the same way, the computation becomes very similar to the computation of \cite{firstpaper}. 
In the next formula we introduce an $S_n[S_2]$ subgroup that belongs to the stabilizer of $C^{4\mu}_J$. 
The reader should recall\cite{firstpaper} that $P_{[A]}$ is defined using the embedding of $S_n[S_2]$ that stabilizes $(1,2)(3,4)\cdots (2n-1,2n)$.
It is now straight forward to obtain
\bea
   \langle O^{\sigma_{4\nu}}_R(Z)\bar{O}^{\sigma_{4\mu}}_S(Z)\rangle
  &=&{\delta_{RS}\, n!2^n \over (2n)!\, d_R} \sum_{\psi_1\in {\cal B}_n}\sum_{\psi_2\in S_n[S_2]} {\rm Tr}(P_{[A]}\Gamma^R(\sigma_{\mu\nu}\psi_1\psi_2))
  C^{4\mu}_I(\psi_1\psi_2)^I_JC^J_{4\mu}\cr
  &=&{\delta_{RS}\, n!2^n \over (2n)!\, d_R} \sum_{\psi_1\in {\cal B}_n}\sum_{\psi_2\in S_n[S_2]} {\rm Tr}(P_{[A]}\Gamma^R(\sigma_{\mu\nu}\psi_1\psi_2))
  C^{4\mu}_I(\psi_1)^I_JC^J_{4\mu}\cr
  &=&{\delta_{RS}\, (n!2^n)^2 \over (2n)!\, d_R} \sum_{\psi_1\in {\cal B}_n} {\rm Tr}(P_{[A]}\Gamma^R(\sigma_{\mu\nu}\psi_1)\hat{P}_{[S]}) 
      C^{4\mu}_I(\psi_1)^I_JC^J_{4\mu}\cr
  &=&{\delta_{RS}\, (n!2^n)^2 \over (2n)!\, d_R}  {\rm Tr}(P_{[A]}\Gamma^R(\sigma_{\mu\nu}) \hat{P}_{[S]})\prod_{i\in {\rm odd\, boxes\, in\,}R} c_i\cr
  &=&\delta_{RS}2^{l(\nu)+l(\mu)}{\chi_{R/4}(\mu)\chi_{R/4}(\nu)\over d_R^2}\prod_{i\in {\rm odd\, boxes\, in\,}R} c_i
  \label{nicetwopoint}
\eea
which is a remarkably simple formula. 
To obtain the last line we have used the results of \cite{ivanov} as explained in Appendix C of \cite{firstpaper}.
Using this in a completely straight forward way we find
\bea
\langle {\rm Tr}(\mu (Z^2)^{\otimes {n\over 2}}) {\rm Tr}(\nu (\bar{Z}^2)^{\otimes {n\over 2}})\rangle =
\sum_{R/4\vdash {n\over 2}}2^{l(\nu)+l(\mu)}\chi_{R/4}(\mu)\chi_{R/4}(\nu)\prod_{i\in {\rm odd\, boxes\, in\,}R} c_i
\label{completeresult}
\eea

\section{A simpler description of the $SO(N)$ basis}

In this section we will argue that the result (\ref{completeresult}) allows us to write a simpler description of our basis,
that is closely related to the Schur polynomial basis of the $U(N)$ theory.
Towards this end, we begin by computing the correlator $\langle O_R \bar{O}_S^{\sigma_{4\nu}}\rangle$.
First, it already follows from the results of \cite{firstpaper} that
\bea
  \langle O_R \bar{O}_S^{\sigma_{4\nu}}\rangle \propto \delta_{RS}
\eea
so that we only need to compute $\langle O_R \bar{O}_R^{\sigma_{4\nu}}\rangle$. Now, noting that 
$O_R=O_R^{\sigma_{4\nu}}$ with $\nu =1^{n\over 2}$, we can apply (\ref{nicetwopoint}) to find
\bea
   \langle O_R \bar{O}_S^{\sigma_{4\nu}}\rangle =\delta_{RS}2^{{n\over 2}+l(\mu)}{d_{R/4}\over d_R^2}\chi_{R/4}(\mu)
              \prod_{i\in {\rm odd\, boxes\, in\,}R} c_i
        \label{proj}
\eea
Now, since the $O_R$ constitute a basis and since (\ref{proj}) gives the two point function of $\bar{O}_S^{\sigma_{4\nu}}$
with any $O_R$, it is clear that (\ref{proj}) can be used to determine $\bar{O}_S^{\sigma_{4\nu}}$ as a linear combination 
of the $O_R$.
We find
\bea
   O_R^{4\nu}={2^{l(\nu)-{n\over 2}}\over d_{R/4}}\chi_{R/4}(\nu)O_R
\eea
Thus, we can now write
\bea
  {\rm Tr}_{V^{\otimes\, n}}(\nu (Z^2)^{\otimes {n\over 2}})&=&\sum_{R/4\vdash {n\over 2}} d_R O_R^{\sigma_{4\nu}}\cr
              &=& \sum_{R/4\vdash {n\over 2}}{d_R\over d_{R/4}}2^{l(\nu)-{n\over 2}}\chi_{R/4}(\nu)O_R
\eea
Notice that, using character orthogonality, we can now invert this relation. Indeed
\bea
\sum_{\nu\in S_{n\over 2}}2^{-l(\nu)}\chi_{S/4}(\nu){\rm Tr}_{V^{\otimes\, n}}(\nu (Z^2)^{\otimes {n\over 2}})
&=& \sum_{R/4\vdash {n\over 2}}{d_R\over d_{R/4}}2^{-{n\over 2}}\Big[\sum_{\nu\in S_{n\over 2}}\chi_{S/4}(\nu)\chi_{R/4}(\nu)\Big]O_R\cr
&=& \sum_{R/4\vdash {n\over 2}}{d_R\over d_{R/4}}2^{-{n\over 2}}\Big[\left({n\over 2}\right)! \delta_{RS}\Big] O_R\cr
&=& {d_S\over d_{S/4}}2^{-{n\over 2}}\Big[\left({n\over 2}\right)! \Big] O_S
\eea
Consequently
\bea
  O_S (Z) ={d_{S/4}\over d_{S}}2^{{n\over 2}} {1\over \left({n\over 2}\right)!}
           \sum_{\nu\in S_{n\over 2}}2^{-l(\nu)}\chi_{S/4}(\nu){\rm Tr}_{V^{\otimes\, n}}(\nu (Z^2)^{\otimes {n\over 2}})
  \label{niceoperatorformula}
\eea
The normalization in (\ref{niceoperatorformula}) looks rather unnatural. From now on we will adopt a new normalization,
given by
\bea
  \chi^0_S (Z)= {1\over \left({n\over 2}\right)!}
      \sum_{\nu\in S_{n\over 2}}2^{-l(\nu)}\chi_{S/4}(\nu){\rm Tr}_{V^{\otimes\, n}}(\nu (Z^2)^{\otimes {n\over 2}})
  \label{niceoperators}
\eea
Notice that we continue to label our operators by $S$, not by $S/4$. 
This deserves a few comments.
In the context of $SO(N)$ gauge theory, the Wick contractions are a sum over elements of $S_{2n}$.
To construct our operators, we have constructed projectors\cite{firstpaper} using representations $S$ of $S_{2n}$. 
Orthogonality of our operators then follows as a consequence of the fact that these projectors commute with the Wick
contractions and are mutually orthogonal.
The two point function is given in terms of a product of factors of boxes in $S$, which were obtained \cite{firstpaper}
by evaluating the action of Jucys-Murphy elements on states in the carrier space of $S$. 
Clearly, $S$ summarizes information about the group theory used to construct our operators: it is the representation 
that organizes the $2n$ indices of the $Z^{ij}$ fields that appear in $O_S$.
Thus for example, cut offs due to the stringy exclusion principle \cite{Maldacena:1998bw,mst} cut $S$ off at
$N$ rows.
$S/4$ on the other hand, plays the role of a useful auxiliary label.
Indeed, $S/4$ is particularly useful in enumerating the possible operators 
(all possible $S/4$ are allowed; only special representations $S$ are allowed) in our basis.
Note also that for the operators $Q_R(Z)$, since the rows in $R$ all have an odd length, there is no notion of the $R/4$ 
label as defined above.
With this new normalization we obtain the lovely answer
\bea
   \langle \chi^0_R (Z) \chi^0_S (\bar{Z})\rangle =\delta_{RS}\prod_{i\in {\rm odd\, boxes\, in\,}S} c_i\equiv \delta_{RS}f_{R,{\rm odd}}
   \label{twopointsfornew}
\eea
The formulas of this section indicate a very interesting interplay between Young diagrams $S\vdash 2n$ and $S/4\vdash {n\over 2}$.

One immediate application of the new formula (\ref{niceoperators}) is in the derivation of a product rule. Indeed,
for $S\vdash 2m$ and $R\vdash 2n$ we have
\bea
     \chi^0_S(Z)\chi^0_R(Z) &=& {1\over \left({n\over 2}\right)!\left({m\over 2}\right)!}
      \sum_{\nu\in S_{m\over 2}} \sum_{\mu\in S_{n\over 2}}
      2^{-l(\nu)-l(\mu)}\chi_{S/4}(\nu)\chi_{R/4}(\mu)
      {\rm Tr}_{V^{\otimes\, n}}(\nu (Z^2)^{\otimes {n\over 2}})
      {\rm Tr}_{V^{\otimes\, n}}(\mu (Z^2)^{\otimes {n\over 2}})\cr
      &=& {1\over \left({n\over 2}\right)!\left({m\over 2}\right)!}
      \sum_{\nu\in S_{m\over 2}} \sum_{\mu\in S_{n\over 2}}\sum_{\sigma\in S_{n+m\over 2}}
      2^{-l(\nu)-l(\mu)}\chi_{S/4}(\nu)\chi_{R/4}(\mu)\delta (\sigma \mu^{-1}\circ\nu^{-1})\cr
      &&\quad {\rm Tr}_{V^{\otimes\, n+m}}(\sigma (Z^2)^{\otimes {n\over 2}})\cr
     &=& {1\over \left({n\over 2}\right)!\left({m\over 2}\right)!\left({n+m\over 2}\right)!}
      \sum_{\nu\in S_{m\over 2}} \sum_{\mu\in S_{n\over 2}}\sum_{\sigma\in S_{n+m\over 2}}\sum_{T/4\vdash {n+m\over 2}}
      2^{-l(\sigma)}\chi_{S/4}(\nu)\chi_{R/4}(\mu)\chi_T(\mu^{-1}\circ\nu^{-1})\cr
      &&\quad \chi_T(\sigma){\rm Tr}_{V^{\otimes\, n+m}}(\sigma (Z^2)^{\otimes {n+m\over 2}})\cr
      &=& \sum_{T/4\vdash {n+m\over 2}}g_{R/4\,S/4\,T/4} \chi^0_T(Z)
\eea
where we have used the formula\cite{FH}
\bea
g_{R/4\,S/4\,T/4}={1\over \left({n\over 2}\right)!\left({m\over 2}\right)!}
      \sum_{\nu\in S_{m\over 2}} \sum_{\mu\in S_{n\over 2}}
      \chi_{S/4}(\nu)\chi_{R/4}(\mu)\chi_T(\mu^{-1}\circ\nu^{-1})
\eea
for the Littlewood-Richardson coefficient.

\section{Multi trace Correlators Again}

The answer (\ref{completeresult}) gives a complete description of correlators in the trace basis. 
To obtain explicit answers we need to evaluate symmetric group characters. 
We explain how this evaluation is carried out in this section. 

In the $U(N)$ case, very similar formulas have been obtained, exploiting the relation between operators 
written in the trace basis and the Schur polynomials.
We could also follow this route given the new form of our operators in (\ref{niceoperators}).
Following this route, we would write multi-point correlators in terms of products of our operators and 
then evaluate these products using the Littlewood-Richardson coefficients.
This approach computes the general multi-trace operators knowing nothing more than a character for an $n$ cycle of
$S_n$ in a Young diagram labeled by a hook representation and the Littlewood-Richardson coefficients.

For $SO(N)$ we can employ the formula (\ref{completeresult}) which requires the computation of characters, beyond the 
character for an $n$ cycle of $S_n$ in a Young diagram labeled by a hook representation. 
As we explain below, the evaluation of these characters is straight forward. 
Given the values of the characters we obtain below, the current computation could also be used to give an alternative 
derivation of the known $U(N)$ correlation functions.

To start, consider the computation for correlators of the form
\bea
   \langle {\rm Tr} (Z^{2J_1}){\rm Tr} (Z^{2J_2}){\rm Tr} (\bar{Z}^{2J_3})\rangle
    =2^3\sum_{R/4\vdash J_3}\chi_{R/4}(\mu)\chi_{R/4}(\nu)\prod_{i\in {\rm odd\, boxes\, in\,}R} c_i
\eea
We have $J_3=J_1+J_2$ and $\mu$ is a $J_3$ cycle while $\nu$ is the product of a $J_1$ cycle and a $J_2$ cycle.
We know that $\chi_{R/4}(\mu)$ will only be non-zero when $R/4$ is a hook. Thus, we will consider only Young diagrams
$R/4$ of the form\footnote{This notation for the Young diagram is listing row lengths, i.e. $[k,1^{J_3-k}]$ has
$k$ boxes in its first row and then it has $J_3-k$ rows which each have a single box.} $[k,1^{J_3-k}]$. A formula
that will be useful is
\bea
\prod_{i\in {\rm odd\, boxes\, in\,}R{\rm \, with\,}R/4=[k,1^{J_3-k}]} c_i = {(N+2k-2)!\over (N- 2J_3 + 2k -2)!}
\eea
We know that 
\bea
   \chi_{[k,1^{J_3-k}]}(\mu)=(-1)^{J_3-k}
\eea
Now consider $\chi_{[k,1^{J_3-k}]}(\nu)$. According to the Murnaghan-Nakayama rule\cite{FH}, this character is equal to a sum
over all ways of extracting a border strip tableau of length $J_1$. For each possible extraction we have to multiply
by $(-1)^h$ where $h$ is the height (= number of rows) of the removed border strip multiplied by the character
of a $J_2$ cycle in the irreducible representation labeled by the Young diagram obtained by removing the border strip from $[k,1^{J_3-k}]$.
Thus, (in the following $(J_2)$ denotes a $J_2$ cycle)
\bea
   \chi_{[k,1^{J_3-k}]}(\nu) &=& -\theta (k>J_1)\chi_{[k-J_1,1^{J_3-k}]}((J_2))+(-1)^{J_1}\theta (J_3-k\ge J_1)\chi_{[k,1^{J_3-k-J_1}]}((J_2))\cr
                             &=& -\theta (k>J_1)\chi_{[k-J_1,1^{J_3-k}]}((J_2))+(-1)^{J_1} \theta (J_2\ge k)\chi_{[k,1^{J_2-k}]}((J_2))\cr
                             &=&  \theta (k>J_1)(-1)^{J_3-k+1}+\theta (J_2\ge k)(-1)^{J_2+J_1-k}\cr
                             &=&  \theta (k>J_1)(-1)^{J_3-k+1}+\theta (J_2\ge k)(-1)^{J_3-k}
\eea
where
\bea
    \theta (k>J_1) &=& 1\quad {\rm if}\quad k > J_1 \cr
                   &=& 0\quad {\rm otherwise}
\eea
\bea
    \theta (J_3-k\ge J_1) &=& 1\quad {\rm if}\quad J_3-k\ge J_1 \cr
                          &=& 0\quad {\rm otherwise}
\eea
It is now clear that
\bea
   \chi_{[k,1^{J_3-k}]}(\mu)\chi_{[k,1^{J_3-k}]}(\nu)=\theta (J_2\ge k)-\theta (k>J_1)
   \label{nicecharacterformula}
\eea
Below we will want to generalize this formula a bit. Towards this end it is worth looking back and realizing that
the negative sign above arose because we removed $J_1$ boxes from the first row - giving a removed tableaux with
height 1. Bear this in mind when considering the subsequent character formulas we obtain, since we will again obtain 
$\theta$ functions with a sign determined by how many border strip tableau were removed from the first row.

Using this character formula we have
\bea
   \langle {\rm Tr} (Z^{2J_1}){\rm Tr} (Z^{2J_2}){\rm Tr} (\bar{Z}^{2J_3})\rangle
    &=&2^3\sum_{R/4\vdash J_3}\chi_{R/4}(\mu)\chi_{R/4}(\nu)\prod_{i\in {\rm odd\, boxes\, in\,}R} c_i\cr
    &=&2^3\sum_{k=1}^{J_3}(\theta (J_2\ge k)-\theta (k>J_1)){(N+2k-2)!\over (N - 2J_3 + 2k -2)!}\cr
    &=&8\left(\sum_{k=1}^{J_2}-\sum_{k=J_1+1}^{J_3}\right){(N+2k-2)!\over (N - 2J_3 + 2k -2)!}
\eea

We now consider correlators of four traces. 
Again, use (\ref{completeresult}) to obtain
\bea
   \langle {\rm Tr} (Z^{2J_1}){\rm Tr} (Z^{2J_2}){\rm Tr} (Z^{2J_3}){\rm Tr} (\bar{Z}^{2J_4})\rangle
    =2^4\sum_{R/4\vdash J_4}\chi_{R/4}(\mu)\chi_{R/4}(\nu)\prod_{i\in {\rm odd\, boxes\, in\,}R} c_i
\eea
We have $J_4=J_1+J_2+J_3$ and $\mu$ is a $J_4$ cycle while $\nu$ is the product of a $J_1$ cycle, a $J_2$ cycle and a $J_3$ cycle.
We know that $\chi_{R/4}(\mu)$ will only be non-zero when $R/4$ is a hook. Thus, we will consider only Young diagrams
$R/4$ of the form $[k,1^{J_4-k}]$. Arguing exactly as we did above, a simple application of the Murnaghan-Nakayama rule gives
\bea
   \chi_{[k,1^{J_4-k}]}(\mu)\chi_{[k,1^{J_3-k}]}(\nu)=&&\theta(k\le J_1)-\theta (J_2< k\le J_1+J_2)\cr
                                                      &&-\theta(J_3< k\le J_1+J_3)+\theta (J_2+J_3<k\le J_4)
\eea
The first term on the right hand side comes from removing both cycles $(J_2)$ and $(J_3)$ from the column.
The second term on the right hand side comes from removing cycle $(J_2)$ from row 1 and cycle $(J_3)$ from the column.
The third term on the right hand side comes from removing cycle $(J_3)$ from row 1 and cycle $(J_2)$ from the column.
The fourth term on the right hand side comes from removing both cycles $(J_2)$ and $(J_3)$ from the first row.
Thus
\bea
   \langle {\rm Tr} (Z^{2J_1})&&\!\!\!\!\!\!\!\!\! {\rm Tr} (Z^{2J_2}){\rm Tr} (Z^{2J_3}){\rm Tr} (\bar{Z}^{2J_4})\rangle
    =2^4\sum_{R/4\vdash J_4}\chi_{R/4}(\mu)\chi_{R/4}(\nu)\prod_{i\in {\rm odd\, boxes\, in\,}R} c_i\cr
    &&=2^4\sum_{k=1}^{J_4}(\theta (k\le J_1)-\theta (J_2< k\le J_1+J_2)-\theta (J_3< k\le J_1+J_3)\cr
    &&\,\,\, +\theta (J_2+J_3+1<k\le J_4)){(N+2k-2)!\over (N - 2J_4 + 2k -2)!}\cr
    &&=16\left(\sum_{k=1}^{J_1}-\sum_{k=J_2+1}^{J_1+J_2}-\sum_{k=J_3+1}^{J_1+J_3}+\sum_{k=J_2+J_3+1}^{J_4}
\right){(N+2k-2)!\over (N - 2J_4 + 2k -2)!}
\eea

Now that we have understood how the signs in the character come about upon applying the Murnaghan-Nakayama rule, 
with a little thought we can obtain the general correlator ($J_n=J_1+J_2+\cdots+J_{n-1}$)
\bea
\langle \prod_{i=1}^{n-1}{\rm Tr} (Z^{2J_i}) {\rm Tr} (\bar{Z}^{2J_n})\rangle &=&
2^n\left(
\sum_{k=1}^{J_1}-\sum_{k=J_2+1}^{J_1+J_2}-\cdots -\sum_{k=J_{n-1}+1}^{J_1+J_{n-1}}+
\sum_{k=J_2+J_3+1}^{J_1+J_2+J_3}+\cdots+\sum_{k=J_2+\cdots +J_{n-1}}^{J_n}
\right)\cr
&&\times {(N+2k-2)!\over (N - 2J_n + 2k -2)!}
\eea
This provides a complete generalization of the $U(N)$ result which was derived in \cite{Caputa:2012dg}.
See also\cite{Kristjansen:2002bb,Beisert:2002bb,Corley:2002mj,Okuyama:2002zn}.

There is a particularly interesting double scaling limit of $\mathcal{N}=4$ super Yang-Mills 
theory\cite{Berenstein:2002jq,Kristjansen:2002bb} defined by
\begin{equation}
N\to\infty~~\text{and}~~J\to\infty~~\text{with}~~\frac{J^2}{N}~~\text{fixed},~~g_{YM}~~\text{fixed}\label{DS}
\end{equation}
where $J$ is the number of fields in the trace. In this limit some non-planar diagrams survive, leading to a new renormalized
genus counting parameter ${J^2\over N}$. This limit is AdS/CFT dual to a pp-wave limit of $AdS_5\times S^5$, in which the superstring 
theory can be quantized. Since we have computed two and three point correlators we can explore this limit in the $SO(N)$ gauge theory.

In order to extract the double scaling limit of the two and three point correlators we will need the following identity \cite{Dhar:2005su}
\begin{eqnarray}
\frac{\G(N+p_1+1)}{\G(N-p_0)}&=&\frac{(N+p_1)!}{(N-p_0-1)!}=\prod^{p_0+p_1}_{l=0}(N+p_1-l)\nn\\
&=&N^{p_0+p_1+1}\exp\left[\sum^{p_0+p_1}_{l=0}\ln(1+\frac{p_1-l}{N})\right].
\end{eqnarray}
Expanding for large $N$ and summing over $l$ yields
\begin{eqnarray}
\frac{\G(N+p_1+1)}{\G(N-p_0)}\sim N^{p_0+p_1+1}\exp\left[\frac{1}{2N}(p_1-p_0)(p_0+p_1+1)+O(1/N^2)\right].
\end{eqnarray}
Applying it to our SO(N) correlators gives
\begin{eqnarray}
\langle\Tr(\bar{Z}^{2J})Tr(Z^{2J})\rangle=4\sum^{J}_{k=1}\frac{\G(N+2(J-k)+1)}{\G(N-2k+1)}\sim\nn\\
 4N^{2J}\sum^{J}_{k=1}\exp\left(\frac{J(2J-4k+1)}{N}\right)
\sim 4J\,N^{2J}\,\frac{\sinh\frac{2J^2}{N}}{\frac{2J^2}{N}}.
\end{eqnarray}
and
\begin{eqnarray}
\langle\Tr(\bar{Z}^{2J_3})\Tr(Z^{2J_1})\Tr(Z^{2J_2})\rangle=8\left(\sum^{J_1}_{k=1}-\sum^{J_3}_{k=J_2+1}\right)\frac{\G(N+2(J_3-k)+1)}{\G(N-2k+1)}\nn\\
\sim 8\times 4N^{2J_3}e^{\frac{J_3}{N}}\frac{\sinh\frac{2J_1J_3}{N}\sinh\frac{2J_2J_3}{N}}{e^{\frac{4J_3}{N}}-1}
\sim 4\times 2J_3N^{2J_3}\frac{\sinh\frac{2J_1J_3}{N}\sinh\frac{2J_2J_3}{N}}{\frac{J^2_3}{N}}
\end{eqnarray}
Compare these results with U(N) correlators in the double scaling limit \cite{Kristjansen:2002bb}
\begin{eqnarray}
\langle\Tr(\bar{Z}^{2J})\Tr(Z^{2J})\rangle=\sum^{2J}_{k=1}\frac{\G(N+k)}{\G(N-2J+k)}
\sim 2J\,N^{2J}\,\frac{\sinh\frac{2J^2}{N}}{\frac{2J^2}{N}}
\end{eqnarray}
\begin{eqnarray}
\langle\Tr(\bar{Z}^{2J_3})\Tr(Z^{2J_1})\Tr(Z^{2J_2})\rangle=\left(\sum^{2J_3}_{k=2J_2+1}-\sum^{2J_1}_{k=1}\right)\frac{\G(N+k)}{\G(N-2J_3+k)}\nn\\
\sim 2N^{2J_3}e^{\frac{2J_3}{N}}(\coth\frac{J_3}{N}-1)\sinh\frac{2J_1 J_3}{N}\sinh\frac{2J_2 J_3}{N}\nn\\
\sim2J_3 N^{2J_3}\frac{\sinh\frac{2J_1J_3}{N}\sinh\frac{2J_2J_3}{N}}{\frac{J^2_3}{N}}
\end{eqnarray}
It is clear that non-planar unoriented diagrams in $SO(N)$ gauge theory do not survive this limit.

Our result is similar to earlier results obtained for the double scaling limit of the matrix model relevant for the $c=1$ string\cite{Brezin:rb}.
From this point of view, the ribbon graphs of the matrix model are identified as a triangulation of the string worldsheet.
These double scaling limits take $N\to\infty$ simultaneously with the world sheet continuum limit in such a way that the 
string coupling is held finite, so that sums over continuum surfaces of any topology are captured.
This limit for antisymmetric matrices has been discussed in \cite{Bilal:1990yy}. 
In this double scaling regime too, only orientable surfaces survive.

\section{Counting}

When constructing gauge invariant operators, it is possible to contract indices using any invariant tensors.
For $SO(N)$ there are two such tensors: the Kronecker delta (which contracts pairs of indices) and the
$\epsilon^{i_1 i_2\cdots i_N}$ tensor. 
In \cite{firstpaper}, taking both invariant tensors into account, we made a precise conjecture for the
basis that can be constructed. In this section we would like to count the number of operators we proposed and 
thereby verify that it is indeed a complete set. We will focus on the case that $N$ is even. 
We will end this section with a few comments on the odd $N$ case.

To start, recall the discussion of \cite{firstpaper}: operators of the form (\ref{opone}) are the complete set
of operators that can be built without using $\epsilon^{i_1 i_2\cdots i_N}$. They correspond to the set of operators
that can be written as a product of traces of even powers of $Z$. Further, they are labeled by Young diagrams with 
an even number of boxes in each column and row. As explained in \cite{firstpaper}, these Young diagrams $R$ built using $2n$ 
boxes with the restriction that $R$ has not more than $N$ rows, can be indexed by partitions of ${n\over 2}$ that have
no more than ${N\over 2}$ parts. Consequently, we can write the partition function for the number of operators
of the form (\ref{opone}) as
\bea
  F_1(x)=\prod_{i=1}^{N\over 2}{1\over 1-x^{2i}}
  \label{evenum}
\eea
The coefficient of the $x^p$ in the expansion of $F_1(x)$ tells us how many operators can be constructed using $p$
$Z$ fields. Next we need to count the number of operators (\ref{optwoq}) built using $\epsilon^{i_1 i_2\cdots i_N}$. These operators
all have a dimension $\ge{N\over 2}$. They are constructed by contracting the indices of ${N\over 2}$ $Z$ fields
with $\epsilon^{i_1 i_2\cdots i_N}$ and contracting the remaining indices in pairs. These can also be constructed
in the form (\ref{opone}), with labels $R$ that consist of a single column of $N$ boxes with a second Young
diagram stacked to the right. To get a non-zero operator, this second Young diagram must again have an even number
of columns and rows. Clearly then, the partition function for the operators constructed using one 
$\epsilon^{i_1 i_2\cdots i_N}$ is
\bea
  F_2(x)=x^{N\over 2}F_1(x) = x^{N\over 2}\prod_{i=1}^{N\over 2}{1\over 1-x^{2i}}
  \label{oddnum}
\eea
Operators constructed using any even number of $\epsilon^{i_1 i_2\cdots i_N}$s lead to Young diagrams that have both
an even number of rows and columns, and hence are included in (\ref{evenum}). 
Similarly, operators constructed using any odd number of $\epsilon^{i_1 i_2\cdots i_N}$s are included in (\ref{oddnum}).
Consequently, the complete partition function for the operators in our basis is
\bea
  F(x)=(1+x^{N\over 2})\prod_{i=1}^{N\over 2}{1\over 1-x^{2i}}
  \label{exactpart}
\eea

The partition for free Yang Mills theory on a compact space has been computed in \cite{Aharony:2003sx}.
The result is
\bea
  G(x)=\int [dO]\sum_{n=0}^\infty x^{nE}\chi_{{\rm Sym}^n(R)}(O) = \int [dO] e^{\sum_{m=1}^\infty {x^{mE}\over m}\chi_R(O^m)} 
  \label{YMPF}
\eea
Here we take $R$ to be the adjoint representation (since our field
$Z$ transforms in the adjoint) and ${\rm Sym}^n(R)$ is the representation
obtained by taking the symmetric product of $n$ copies of the adjoint.
Set $E=1$ and then expand to get a polynomial in $x$. 
The coefficient of $x^n$ counts the number of operators that can be built using $n$ $Z$s.
Consequently, if our basis is complete, we should find $G(x)=F(x)$.
To evaluate (\ref{YMPF}) we need the adjoint character (we focus on $SO(2n)$)
\bea
   \chi_R({\rm x}) =\sum_{1\le i<j\le n}(x_i^{}x_j^{}+x_i^{-1}x_j^{}+x_i^{}x_j^{-1}+x_i^{-1}x_j^{-1})+n
\eea
and the integration measure valid for any symmetric $f({\rm x})$, ${\rm x}=(x_1,...,x_n)$\cite{Dolan:2008qi}
\bea
\int_{\rm SO(2n)} [dO] f({\rm x}) ={1\over 2^{n-1}n!}\int_{T_n}\prod_{j=1}^n {dx_j\over 2\pi i x_j}\Delta({\rm x}+{\rm x}^{-1})^2f({\rm x}) 
\eea
Here $T_n = S^1\times S^1\times \cdots \times S^1$ is the unit torus.
Using this we find
\bea
  G(x) &=& \int_{\rm SO(2n)} [dO] e^{\sum_{m=1}^\infty {x^{m}\over m}\chi_R(O^m)}\cr
       &=& {1\over 2^{n-1}n!}\int_{T_n}\prod_{j=1}^n {dx_j\over 2\pi i x_j}\Delta({\rm x}+{\rm x}^{-1})^2\cr
&&\times e^{\sum_{m=1}^\infty {x^{m}\over m} \left(
\sum_{1\le i<j\le n}(x_i^{m}x_j^{m}+x_i^{-m}x_j^{m}+x_i^{m}x_j^{-m}+x_i^{-m}x_j^{-m})+n
\right)}\cr
       &=& {1\over 2^{n-1}n!}\int_{T_n}\prod_{j=1}^n {dx_j\over 2\pi i x_j}\Delta({\rm x}+{\rm x}^{-1})^2 \cr
&&\times \prod_{1\le i<j\le n}{1\over (1-x)^n}{1\over 1-x x_i x_j}{1\over 1-x x_i^{-1}x_j}
                                     {1\over 1-x x_i x_j^{-1}}{1\over 1-x x_i^{-1}x_j^{-1}}
\eea
Some straight forward algebra shows that
\bea
  G(x)= {1\over 2^{n-1}n!}\int_{T_n}\prod_{j=1}^n {dx_j\over 2\pi i x_j}
        \prod_{1\le i<j\le n}{1\over (1-x)^n}F_{ij}
\eea
where
\bea
 F_{ij} = {1- x_i x_j\over 1-x x_i x_j}{1- x_i^{-1}x_j\over 1-x x_i^{-1}x_j}
          {1-x_i x_j^{-1}\over 1-x x_i x_j^{-1}}{1-x_i^{-1}x_j^{-1}\over 1-x x_i^{-1}x_j^{-1}}
\eea
We have not managed to compute these integrals for general $n$. 
We have however verified that (the integrals over $x_i$ below all run over the unit circle; we have taken $x<1$)
\bea
{1\over 4}\int {dx_1\over 2\pi i x_1}\int {dx_2\over 2\pi i x_2} F_{12}={1\over (1+x)^2}
\eea
so that for $N=2n=4$ we have
\bea
G(x)=(1+x^2){1\over 1-x^2}{1\over 1-x^4}
\eea
and
\bea
     {1\over 24}\int {dx_1\over 2\pi i x_1}\int {dx_2\over 2\pi i x_2}\int {dx_3\over 2\pi i x_3} F_{12}F_{13}F_{23}
         ={1\over (1+x)^2}{1\over 1+x^2}{1\over 1+x+x^2}
\eea
so that for $N=2n=6$ we have
\bea
   G(x)=(1+x^3){1\over 1-x^2}{1\over 1-x^4}{1\over 1-x^6}
\eea
This proves that for $N=4,6$, $G(x)=F(x)$ which provides strong support that we have indeed constructed a basis. 
Given the results of this section, we conjecture that the partition function for $SO(N)$, with $N$ even is given by
(\ref{exactpart}). 
Our results also lead us to conjecture that for $SO(N)$, with $N$ odd, the partition function is
\bea
  G(x)=\prod_{i=1}^{N-1\over 2}{1\over 1-x^{2i}}
  \label{exactoddnum}
\eea

\section{Link to Free Fermions}

The dynamics of gauge invariant operators of a single Hermitian matrix model reduces to eigenvalue dynamics\cite{BIPZ}. 
If one restricts to the gauge invariant and purely holomorphic or antiholomorphic observables of complex matrix models, 
one can again reduce to eigenvalue dynamics\cite{Ginibre:1965zz}.
This eigenvalue dynamics can then be mapped to the dynamics of $N$ free fermions in an external potential\cite{cjr}.
For the free complex matrix model, the basis provided by the Schur polynomials has a particularly close relationship
to free fermion dynamics: the Schur polynomials can be mapped to free fermion wave functions\cite{cjr}. 
In this section our goal is to argue that our operators (\ref{niceoperators}) also describe free fermion wave functions.

Before considering the eigenvalue dynamics it is useful to recall a few facts about the wave functions of the single
particle Hamiltonian
\bea
   H=-{\partial\over\partial z}{\partial\over\partial \bar{z}}+z\bar{z}
\eea
which describes a single particle moving in a harmonic oscillator potential, in two dimensions.
The ground state wave function (not normalized) is
\bea
   \psi_0(z,\bar{z}) = e^{-z\bar{z}}
\eea
Further, at any given energy level, the state with largest angular momentum (again, not normalized) is given by
\bea
   \psi_l(z,\bar{z}) = z^le^{-z\bar{z}}
   \label{LLL}
\eea
The parity of this wave function is given by $(-1)^l$ so that (for example) if $l$ is even we have an even parity state.
These wave functions will play a prominent role below.

To start, return to the two point function (\ref{twopointsfornew}) and rewrite it in terms of eigenvalues
\bea
\delta_{RS}f_{R,{\rm odd}} &=& \langle \chi_R^0 (Z) \chi_S^0 (\bar{Z})\rangle\cr
                           &=& \int [dzd\bar{z}] |\Delta (z) |^2 \chi^0_R(z)\chi^0_S(\bar{z})e^{-2\sum_{i=1}^{N\over 2} z_i\bar{z}_i}
\label{sotwpnt}
\eea
where the required Jacobian $\Delta (z)=\prod_{1\le i<j\le {N\over 2}} (z_i^2-z_j^2)$ is discussed in Appendix \ref{jacs}. 
Consequently, if we introduce the wave functions
\bea
  \psi_R(z,\bar{z})= \Delta (z)\chi^0_R(z)e^{-\sum_{i=1}^{N\over 2} z_i\bar{z}_i}
\eea
the two point equation (\ref{sotwpnt}) becomes the statement that these wave functions are orthogonal.
To find the interpretation of these wave functions, we need to unpack some of the details of the expression 
(\ref{niceoperators}). Study operators built from $n$ fields. Recall that $n$ must be even.
We will focus on $SO(N)$ gauge theory with $N$ even. Denote the eigenvalues of $Z$ by $\pm iz_i$.
Introduce the matrix
\bea
  M=\left[
\begin{array}{cccc}
z_1^2  &0      &\cdots &0\cr
0      &z_2^2  &\cdots &0\cr
\vdots &\vdots &\vdots &\vdots\cr
0      &0      &\cdots &z^2_{N\over 2}
\end{array}
\right]
\eea
Using the result (\ref{traceofX}) from Appendix \ref{jacs}, we have
\bea
   \chi^0_S(z)=(-1)^{n\over 2}\chi_{S/4}(M)
\eea
where $\chi_{S/4}(M)$ is nothing but the Schur polynomial.
The Young diagram $S$ can also be taken as a partition of ${n\over 2}$ with parts $s_1,s_2,...,s_{N\over 2}$
equal to the row lengths of $S$.
Recall that the Schur polynomial obeys the identity
\bea
  \chi_S (y_1,y_2,...,y_{N\over 2})=\prod_{1\le i<j\le {N\over 2}}(y_i-y_j)\times{\rm det}\left[
\begin{array}{cccc}
y_1^{s_1+{N\over 2}-1} &y_2^{s_1+{N\over 2}-1} &\cdots &y_{N\over 2}^{s_1+{N\over 2}-1}\cr
y_1^{s_2+{N\over 2}-2} &y_2^{s_2+{N\over 2}-2} &\cdots &y_{N\over 2}^{s_2+{N\over 2}-2}\cr
\vdots                 &\vdots                 &\vdots &\vdots                \cr
y_1^{s_{N\over 2}}     &y_2^{s_{N\over 2}}     &\cdots &y_{N\over 2}^{s_{N\over 2}}
\end{array}
\right]
\eea
Using this identity we find
\bea
  \psi_R(z,\bar{z})=(-1)^{n\over 2}
{\rm det}\left[
\begin{array}{cccc}
z_1^{r_1+N-2}          &z_2^{r_1+N-2} &\cdots  &z_{N\over 2}^{r_1+N-2}\cr
z_1^{r_2+N-4}          &z_2^{r_2+N-4} &\cdots  &z_{N\over 2}^{r_2+N-4}\cr
\vdots                  &\vdots         &\vdots  &\vdots                \cr
z_1^{r_{N\over 2}}     &z_2^{r_{N\over 2}}     &\cdots &z_{N\over 2}^{r_{N\over 2}}
\end{array}
\right]e^{-\sum_{i=1}^{N\over 2} z_i\bar{z}_i}
\label{wfs}
\eea
The interpretation of this wave function is clear. It is the Slater determinant of ${N\over 2}$ single particle
wave functions, and consequently is the wave function of ${N\over 2}$ fermions. These fermions are not interacting
(apart from the usual Fermi statistics) but are moving in an external harmonic oscillator potential. Each particle
has maximum angular momentum for its energy and is in a parity even state (recall that the number of boxes in each
row of $R$ is even). 

Next consider the wave functions $\varphi_R(z,\bar{z})$ that corresponds to the operators $Q_R(Z)$ given in (\ref{optwoq}).
Since the approach to this problem is identical to what we have done above, it should be no surprise that
the wave functions $\varphi_R(z,\bar{z})$ take exactly the same form as (\ref{wfs}). 
The key difference is that now every row has at least one box and all rows have an odd number of boxes.
Thus, we again have ${N\over 2}$ fermions, each particle again has maximum angular 
momentum for its energy, but now each is in a parity odd state. 
This result should not be surprising. 
We can describe these fermion wave functions in terms of a density in phase space, indicating which states are filled.
For some nice examples, see \cite{Mukhi:2005cv}.
The operators $Q_R(Z)$ include a factor of the Pfaffian. 
In terms of the fermion phase space density, the Pfaffian creates a small hole at the origin of phase space,
pushing all fermions up by one level so that each particle lands up moving from a parity even state into a parity odd state.

There is a simple generalization of this result to $SO(N)$ with $N$ odd. The two point function (\ref{twopointsfornew}) becomes
\bea
\delta_{RS}f_{R,{\rm odd}} &=& \langle \chi_R^0 (Z) \chi_S^0 (\bar{Z})\rangle\cr
                           &=& \int [dzd\bar{z}] |\tilde{\Delta} (z) |^2 \chi^0_R(z)\chi^0_S(\bar{z})e^{-2\sum_{i=1}^{N-1\over 2} z_i\bar{z}_i}
\label{sectwpnt}
\eea
where now $\tilde{\Delta} (z)=\prod_{k=1}^{N-1\over 2}z_k\prod_{1\le i<j\le {N-1\over 2}} (z_i^2-z_j^2)$.
This Jacobian is discussed in Appendix \ref{jacs}.  
A very similar discussion shows that the corresponding wave functions are unchanged.
This follows because in the formula (\ref{wfs}) we need to replace $N\to N-1$ and we need to multiply by the extra
factor $\prod_{k=1}^{N-1\over 2}z_k$. 
Multiplication by the extra factor can be accomplished by sending $r_i\to r_i+1$. 
The combined transformation $N\to N-1$ and $r_i\to r_i+1$ leaves (\ref{wfs}) unchanged.
Since $N$ is odd, the fermions are now all in a parity odd state.

The identification of our operators with free fermion wave functions has an immediate application.
There is a class of ${1\over 2}$ -BPS bubbling geometries which arise as orientifolds of type IIB string theory. 
In \cite{Mukhi:2005cv} these geometries have been put into correspondence with free fermions (see also \cite{Lin:2004nb}).
Using this dictionary, we are able to give the operators dual to certain backgrounds with these geometries.
The description of \cite{Mukhi:2005cv} is in terms of free fermions moving in a harmonic oscillator potential on the half-line.
The description we have constructed above is in terms of the holomorphic sector of two-dimensional free fermions (i.e. 2+1 dimensional) 
moving in a spherically symmetric harmonic oscillator potential.
The two are related by truncating the system of two dimensional fermions (analogous to projecting to the lowest Landau level in the quantum Hall effect).
See \cite{cjr,Donos:2005vm} for a clear discussion of the relevant truncation in the context of the $U(N)$ gauge theory.

\section{$Sp(N)$ gauge theory}

There is a close relation between invariants of representations of $SO(N)$ and representations of $Sp(N)$ obtained by exchanging 
symmetrizations and antisymmetrizations and replacing $N$ by $-N$.
In this section we will argue that the finite $N$ physics of $SO(N)$ and $Sp(N)$ gauge theory are related in exactly the
same way. 

\begin{figure}[ht]%
\begin{center}
\includegraphics[width=15cm]{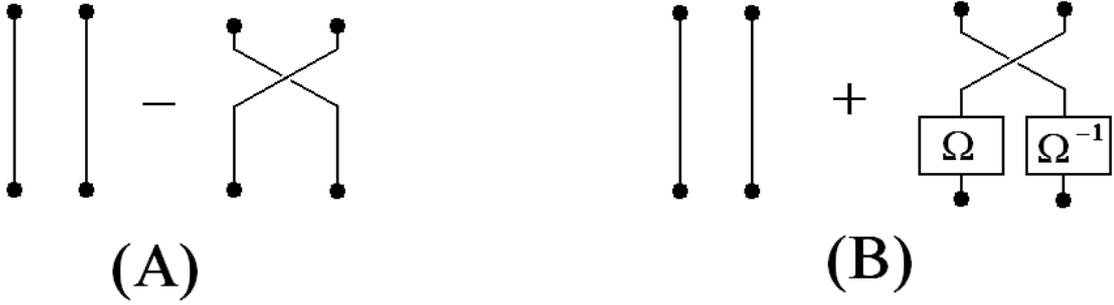}%
\caption{A comparison of the free field two point function of the $SO(N)$ gauge theory (shown in (A) above) and
         of the $Sp(N)$ gauge theory (shown in (B) above).}%
\label{cmpsosp}%
\end{center}
\end{figure}

In $Sp(N)$ gauge theory, we have the following identity obeyed by fields in the adjoint ($X$ is an $N\times N$ matrix with $N$ even)
\be
X^T=-\Omega\, X\, \Omega^T=\Omega\, X\,\Omega
\ee
where the $N\times N$ matrix $\Omega$ is given by ($\mathbf{1}_{d}$ is the $d$ dimensional identity matrix)
\begin{eqnarray}
\Omega=\left(\begin{array}{cc}
0 & \mathbf{1}_{N\over 2}\\
-\mathbf{1}_{N\over 2} & 0\end{array}\right),
\end{eqnarray}
The free field two point function is
\be
\langle X_{ij}\,X_{kl}\rangle =\delta_{il}\delta_{jk}-\Omega_{ik}\Omega_{jl}
\ee
Note that $\Omega$ clearly satisfies the following relations
\begin{equation}
\Omega^2=-\mathbf{1}_N,\qquad \Omega^{-1}=\Omega^{T}=-\Omega.
\end{equation}
A clear consequence is that only traces with an even number of fields survive
\bea
\Tr(X^{J})=\Tr\left((X^J)^{T}\right)=(\Omega^2)^J\Tr(X^J)=(-1)^J\Tr(X^J)
\eea
Now, consider a two matrix model with real fields $X$ and $Y$. 
Combine these real fields into the complex combinations
\bea
   Z^{ab}=X^{ab}+iY^{ab},\qquad Z_{ab}=X^{ba}+iY^{ba}
\eea
The free field $Sp(N)$ two point functions are given by $\langle Z^{ij}Z^{kl}\rangle=0$,
$\langle Z_{ij} Z_{kl}\rangle=0$ and
\bea
   \langle Z^{ij} Z_{kl}\rangle=\delta^{i}_{k}\delta^{j}_{l}+(\Omega)^{i}_{l}(\Omega^{-1})^{j}_{k}
\eea
This looks very similar to the two point functions of the $SO(N)$ gauge theory (see figure \ref{cmpsosp}).
Consider the computation of $\langle T^{\s_{4\n}}_{\s_{4\m}}\rangle$ where
\bea
T^{\s_{4\n}}_{\s_{4\m}} = C^{\s_{4\n}}_J\beta^J_{i_1i_2\cdots i_{2n-1}i_{2n}}Z^{i_1i_2}\cdots Z^{i_{2n-1}i_{2n}}
        C_{\s_{4\m}}^K\tau_K^{l_1 l_2\cdots l_{2n-1}l_{2n}}Z_{l_1l_2}\cdots Z_{l_{2n-1}l_{2n}}
   \label{intermed}
\eea
in the free field theory. 
Our goal is to relate the computation of this correlator in the $Sp(N)$ theory with the computation of the same correlator, but in the $SO(N)$ theory.
The contribution coming from keeping only the first term ($=\delta^{i}_{k}\delta^{j}_{l}$) in each Wick contraction (see figure \ref{cmpsosp}) 
is identical for the two.
This term has the form $N^p$.
The contributions coming when we start to include the second term ($=(\Omega)^{i}_{l}(\Omega^{-1})^{j}_{k}$ for $Sp(N)$ and
$=-\delta^{i}_{l}\delta^{j}_{k}$ for $SO(N)$) in the Wick contractions can differ in sign.
Contributions coming from including any odd number of these second terms will differ in sign between the $Sp(N)$ and $SO(N)$ theories,
and will have the form $cN^{p-2i-1}$ for some integer $i$ and $c$ a constant determined by the number of ribbon graphs summed. 
Contributions coming from including any even number of these second terms will have the same sign in the $Sp(N)$ and $SO(N)$ theories,
and will have the form $cN^{p-2k}$ for some integer $k$.
Thus, when $p$ is even the computation of (\ref{intermed}) in the $Sp(N)$ theory gives the same answer as the answer for the $SO(N)$ theory
after we take $N\to -N$.
When $p$ is odd the answer for the $Sp(N)$ theory follows by taking $N\to -N$ in the answer for the $SO(N)$ theory and multiplying by $-1$.
The power $p$ of the leading term is determined rather simply by the sign of the permutations $\beta$ and $\tau$.
To compute the sign of a permutation decompose it into a product of transpositions. 
This decomposition is not unique. 
The sign of the permutation ${\rm sgn}(\s)=(-1)^m$ where $m$ is the number of transpositions in the product.
${\rm sgn}(\s)$ is well defined, i.e. it does not depend on the specific decomposition of $\s$ into transpositions.
Our final result is
\bea
\langle T^{\s_{4\n}}_{\s_{4\m}}\rangle_{Sp(N)}
={\rm sgn}(\beta){\rm sgn}(\tau)\left[
\langle T^{\s_{4\n}}_{\s_{4\m}} \rangle_{SO(N)}\Big|_{N\to -N}\right]\cr
\label{keyresult}
\eea
This result will play an important role below.

Introduce the operators
\bea
  O_R (Z) ={1\over (2n)!}\sum_{\sigma\in S_{2n}}\chi_R(\sigma )
           \sigma^{ i_1 i_2 i_2 i_1\cdots i_{n-1 }i_n i_n i_{n-1}}_{j_1 j_2 \cdots j_{2n-1} j_{2n}}
            Z^{j_1 j_2}\cdots Z^{j_{2n-1} j_{2n}}
\label{opspone}
\eea
\bea
  \bar{O}_R (Z) ={1\over (2n)!}\sum_{\sigma\in S_{2n}}\chi_R(\sigma )
                 \sigma_{ i_1 i_2 i_2 i_1 \cdots i_{n-1}i_n i_n i_{n-1}}^{j_1 j_2 \cdots j_{2n-1} j_{2n}}
                 Z_{j_1 j_2}\cdots Z_{j_{2n-1} j_{2n}}
\label{opsptwo}
\eea
To compute the two point function of these operators we will be using (\ref{keyresult}).
For this reason we will need to spell out the gauge group we are using to compute the correlator.
We will need one more result from group theory. 
Recall that to get the conjugate (or transpose) $R^c$ of a Young diagram $R$ we need to swap rows and columns.
For example
\bea
  R=\yng(4,2)\qquad R^c=\yng(2,2,1,1)
\eea
The characters of $R$ and $R^c$ are related by
\bea
   \chi_R(\s)={\rm sgn}(\s)\chi_{R^c}(\s)
\eea
From the point of view of the projectors used in defining $O_R$, taking the conjugate of $R$ corresponds to
swapping symmetrization and antisymmetrization of indices.
We are now ready to compute the correlation functions of the $O_R$.
For $N$ even, $R\vdash 2n$ and $S\vdash 2m$ we find
\bea
   \langle O_R (Z)\bar{O}_S (Z)\rangle_{Sp(N)}&&=
{1\over (2n)!}{1\over (2m)!}\sum_{\sigma\in S_{2n}}
\sum_{\tau\in S_{2m}}\chi_R(\sigma)\chi_S(\tau )\cr
&&\!\!\!\!\!\!\!\!\!\!\!\!\!\!\!\!\!\!\!\!\!\!\!\!\!\!\!\!\!\!\!\!\!\!\!\!\!\!\!\!\!\!\!\!\!
\times\langle \sigma^{ i_1 i_2 i_2 i_1\cdots i_{n-1 }i_n i_n i_{n-1}}_{j_1 j_2 \cdots j_{2n-1} j_{2n}}
            Z^{j_1 j_2}\cdots Z^{j_{2n-1} j_{2n}}
        \tau_{ i_1 i_2 i_2 i_1 \cdots i_{m-1}i_m i_m i_{m-1}}^{j_1 j_2 \cdots j_{2m-1} j_{2m}}
                 Z_{j_1 j_2}\cdots Z_{j_{2m-1} j_{2m}}\rangle_{Sp(N)}\cr
&&={1\over (2n)!}{1\over (2m)!}\sum_{\sigma\in S_{2n}}
\sum_{\tau\in S_{2m}}\chi_R(\sigma)\chi_S(\tau ){\rm sgn}(\sigma){\rm sgn}(\tau)\cr
&&\!\!\!\!\!\!\!\!\!\!\!\!\!\!\!\!\!\!\!\!\!\!\!\!\!\!\!\!\!\!\!\!\!\!\!\!\!\!\!\!\!\!\!\!\!
\times\langle \sigma^{ i_1 i_2 i_2 i_1\cdots i_{n-1 }i_n i_n i_{n-1}}_{j_1 j_2 \cdots j_{2n-1} j_{2n}}
            Z^{j_1 j_2}Z^{j_3 j_4}\cdots Z^{j_{2n-1} j_{2n}}
        \tau_{ i_1 i_2 i_2 i_1 \cdots i_{m-1}i_m i_m i_{m-1}}^{j_1 j_2 \cdots j_{2m-1} j_{2m}}
                 Z_{j_1 j_2}\cdots Z_{j_{2m-1} j_{2m}}\rangle_{SO(N)}\Big|_{N\to -N}\cr
&&={1\over (2n)!}{1\over (2m)!}\sum_{\sigma\in S_{2n}}
\sum_{\tau\in S_{2m}}\chi_{R^c}(\sigma )\chi_{S^c}(\tau )\cr
&&\!\!\!\!\!\!\!\!\!\!\!\!\!\!\!\!\!\!\!\!\!\!\!\!\!\!\!\!\!\!\!\!\!\!\!\!\!\!\!\!\!\!\!\!\!
\times\langle \sigma^{ i_1 i_2 i_2 i_1\cdots i_{n-1 }i_n i_n i_{n-1}}_{j_1 j_2 \cdots j_{2n-1} j_{2n}}
            Z^{j_1 j_2}\cdots Z^{j_{2n-1} j_{2n}}
        \tau_{ i_1 i_2 i_2 i_1 \cdots i_{m-1}i_m i_m i_{m-1}}^{j_1 j_2 \cdots j_{2m-1} j_{2m}}
                 Z_{j_1j_2}\cdots Z_{j_{2m-1} j_{2m}}\rangle_{SO(N)}\Big|_{N\to -N}\cr
&&=\big[\langle O_{R^c} (Z)\bar{O}_{S^c} (Z)\rangle_{SO(N)}\Big]_{N\to -N}
\eea
Using the results of \cite{firstpaper} we now immediately find\footnote{Recall that as symmetric group representations we have
$d_R=d_{R^c}$. We have used this when we wrote (\ref{mainspresult}).}
\bea
\langle O_R(Z)\bar{O}_S(Z)\rangle_{Sp(N)}
=\delta_{RS}\, 2^{n}\left( {d_{R/4} \over  d_R}\right)^2\,\prod_{i\in {\rm odd\,\, boxes\,\, in\,\,}R^c}c_i\Big|_{N\to -N}
\label{mainspresult}
\eea
It is straight forward to verify that
\bea
  \prod_{i\in {\rm odd\,\, boxes\,\, in\,\,}R^c}c_i\Big|_{N\to -N} = \prod_{i\in {\rm even\,\, boxes\,\, in\,\,}R}c_i
\eea
so that we finally obtain
\bea
\langle O_R(Z)\bar{O}_S(Z)\rangle_{Sp(N)}
=\delta_{RS}\, 2^{n}\left( {d_{R/4} \over  d_R}\right)^2\,\prod_{i\in {\rm even\,\, boxes\,\, in\,\,}R}c_i
\label{finalmainspresult}
\eea
Recall that the even boxes will occupy every second row, including the top most row. As an example we have
filled the even boxes below
\bea
  \young(*****,\,\,\,\,\,,***,\,\,\,)
\eea

We can now repeat many of the arguments we developed for the $SO(N)$ theory.
Since many of the details are basically the same, we will for the most part simply quote the results.
Correlations functions are given by
\be
  \langle\Tr\left(\mu(\bar{Z}^2)^{\otimes \frac{n}{2}}\right)\Tr\left(\nu(Z^2)^{\otimes \frac{n}{2}}\right)\rangle_{Sp(N)}
  =2^{l(\mu)+l(\nu)}\sum_{R/4\vdash n/2}\chi_{R/4}(\mu)\chi_{R/4}(\nu)\prod_{i\in \text{even in R}}c_i
  \label{spncorrelators}
\ee
Using this result we can again give a simpler form for our operators
\bea
  \chi_S^0 (Z)= {1\over \left({n\over 2}\right)!}
      \sum_{\nu\in S_{n\over 2}}2^{-l(\nu)}\chi_{S/4}(\nu){\rm Tr}_{V^{\otimes\, n}}(\nu (Z^2)^{\otimes {n\over 2}})
\eea
The two point function of these operators is
\bea
   \langle \chi_R^0 (Z) \chi_S^0 (\bar{Z})\rangle_{Sp(N)} =\delta_{RS}\prod_{i\in {\rm even\, boxes\, in\,}S} c_i
   \label{sptwopoint}
\eea
These operators again enjoy a simple product rule ($S/4\vdash {n_1\over 2}$, $R/4\vdash {n_2\over 2}$)
\bea
     \chi_S^0(Z)\chi_R^0(Z) = \sum_{T/4\vdash {n_1+n_2\over 2}} g_{R/4\,S/4\,T/4} \chi_T^0(Z)
\eea
where $g_{R/4\,S/4\,T/4}$ is the Littlewood-Richardson coefficient. 

Using (\ref{spncorrelators}) we can again give a formula for general extremal correlation functions
\bea
\langle\Tr(\bar{Z}^{2J_n})\Tr(Z^{2J_1})...\Tr(Z^{2J_{n-1}})\rangle_{Sp(N)}=\nn\\
(-1)^{n}2^n\left(\sum^{J_1}_{k=1}-...+...-...+\sum^{J_n}_{k=J_2+...J_{n-1}+1}\right)\frac{\G(N+2k)}{\G(N-2J_n-2k)}
\eea
where $J_n=\sum^{n-1}_{i=1}J_i$. In particular, for two point functions we find
\be
\langle\Tr\left(\bar{Z}^{2J}\right)\Tr\left(Z^{2J}\right)\rangle_{Sp(N)}=4\sum^{J}_{k=1}\frac{\G(N+2k)}{\G(N-2J-2k)}
\ee
We have also considered the double scaling limit of the two-point functions in the $Sp(N)$ gauge theory. 
We find
\bea
\langle\Tr\left(\bar{Z}^{2J}\right)\Tr\left(Z^{2J}\right)\rangle\sim 4N^{2J}\sum^{J}_{k=1}\exp\left[\frac{J}{N}(4k-2J-1)\right]\nn\\
=4N^{2J}\frac{e^{\frac{J(2J+3)}{N}}-e^{-\frac{J(2J-3)}{N}}}{e^{\frac{4J}{N}}-1}\sim 4JN^{2J}\frac{\sinh\frac{2J^2}{N}}{\frac{2J^2}{N}}
\eea
Similarly for three-point functions
\bea
\langle\Tr(\bar{Z}^{2J_3})\Tr(Z^{2J_1})\Tr(Z^{2J_2})\rangle=-8\left(\sum^{J_1}_{k=1}-\sum^{J_3}_{k=J_2+1}\right) \frac{\G(N+2k)}{\G(N-2(J_3-k))}\nn\\
\sim\,\frac{32N^{2J_3}}{e^{\frac{4J_3}{N}}-1}e^{\frac{3J_3}{N}}\sinh\frac{2J_1J_3}{N}\sinh\frac{2J_2J_3}{N}\sim 8J_3N^{2J_3}\frac{\sinh\frac{2J_1J_3}{N}\sinh\frac{2J_2J_3}{N}}{J^2_3/N}
\eea
Since this is the same as the $SO(N)$ result, we again see that in this double scaling limit only orientable higher genus surfaces contribute. 

Our operators can again be related to free fermion wave functions. 
Indeed, start from
\bea
\delta_{RS}f_{R,{\rm even}} &=& \langle \chi_R^0 (Z) \chi_S^0 (\bar{Z})\rangle\cr
                           &=& \int [dzd\bar{z}] |\Delta (z) |^2 \chi^0_R(z)\chi^0_S(\bar{z})e^{-2\sum_{i=1}^{N\over 2} z_i\bar{z}_i}
\label{sptwpnt}
\eea
where $\Delta (z)=\prod_{k=1}^{N\over 2}z_k\prod_{1\le i<j\le {N\over 2}} (z_i^2-z_j^2)$. 
Consequently,
\bea
  \psi_R(z,\bar{z})= \Delta (z)\chi^0_R(z)e^{-\sum_{i=1}^{N\over 2} z_i\bar{z}_i}
\eea
Arguing as we did for the $SO(N)$ theory, we find
\bea
  \psi_R(z,\bar{z})=
{\rm det}\left[
\begin{array}{cccc}
z_1^{r_1+N-1}          &z_2^{r_1+N-1} &\cdots  &z_{N\over 2}^{r_1+N-1}\cr
z_1^{r_2+N-3}          &z_2^{r_2+N-3} &\cdots  &z_{N\over 2}^{r_2+N-3}\cr
\vdots                  &\vdots         &\vdots  &\vdots                \cr
z_1^{r_{N\over 2}+1}     &z_2^{r_{N\over 2}+1}     &\cdots &z_{N\over 2}^{r_{N\over 2}+1}
\end{array}
\right]e^{-\sum_{i=1}^{N\over 2} z_i\bar{z}_i}
\label{spwfs}
\eea
This is the Slater determinant of ${N\over 2}$ single particle wave functions i.e. the wave function of ${N\over 2}$ fermions. 
These fermions are not interacting but are moving in an external harmonic oscillator potential. 
Each particle has maximum angular momentum for its energy and is in a parity odd state (recall that the number of boxes in each
row of $R$ is even and $N$ is even). This completes our discussion of the $Sp(N)$ gauge theory.

\section{Discussion}

There are a number of interesting questions that could be pursued at this point. 
In this section we will list some of them.

Using the results of this article, it is possible to study the three-point function of 
two giant gravitons and one point like graviton in the $SO(N)$ gauge theory. 
These correlators can also be computed using methods of semiclassical string theory,
using a Born-Infeld description of the giant graviton. 
Computations of this type\cite{Bissi:2011dc,Caputa:2012yj,Hirano:2012vz,Lin:2012ey}
show a perfect match between correlators computed in the $U(N)$ gauge theory
and three point functions computed using semiclassical string theory. 
Computations along these lines will further test the gauge theory results we have obtained.

It has been possible to construct a very detailed holographic map between ${1\over 2}$-BPS
bubbling geometries\cite{Lin:2004nb} and ${\cal N}=4$ super Yang-Mills theory with $U(N)$
gauge group\cite{Skenderis:2007yb}. The connection between the ${1\over 2}$-BPS sector of
${\cal N}=4$ super Yang-Mills theory and the dynamics of free fermions\cite{cjr,Berenstein:2004kk} 
was a key insight that determined much of the holographic dictionary. Given the close
connection to free fermions developed above, it is natural to develop the holography
of bubbling orientifolds of type IIB string theory.

Finally, it would be natural to study the spectrum of anomalous dimensions in non-planar large $N$ limits 
of the $SO(N)$ and $Sp(N)$ gauge theories. 
New integrable sectors of the $U(N)$ gauge theory have been found in these limits\cite{gs,Koch:2011hb,DCI}.
Is integrability also present in large $N$ but non-planar limits of $SO(N)$ and $Sp(N)$ gauge theory?

We hope to return to these issues.

\noindent
{\it Acknowledgements:}
This work is based upon research supported by the South African Research Chairs
Initiative of the Department of Science and Technology and National Research Foundation.
Any opinion, findings and conclusions or recommendations expressed in this material
are those of the authors and therefore the NRF and DST do not accept any liability
with regard thereto.
RdMK would like to thank Collingwood College, Durham for their support.
PC would like to thank Vikram Vyas for hospitality during the last stages of the project.
Finally, the work of PD is supported in part by a Claude Leon Fellowship.

\begin{appendix}

\section{Simplifying the $SO(N)$ basis}\label{detailed}

In this Appendix we will give an alternative derivation of (\ref{niceoperatorformula}). 
Our starting point is
\begin{equation}
O_R^{\sigma_{4\mu}}=\frac{1}{2n!}\sum_{\beta \in S_{2n}}\chi_R(\beta) C_{j_1 j_2 \cdots j_{2n}}^{\sigma_{4\mu}}\big{(}\beta\big{)}_{i_1 i_2\cdots i_{2n}}^{j_1 j_2\cdots j_{2n}}Z^{i_1 i_2}\cdots Z^{i_{2n-1} i_{2n}}
\label{FOP}
\end{equation}
Recall our shorthand notation for indices
\begin{equation}
 C_J^{\sigma_{4\mu}}=C_{j_1 j_2 \cdots j_{2n}}^{\sigma_{4\mu}}, \qquad \big{(}\beta\big{)}_I^J=\big{(}\beta\big{)}_{i_1 i_2\cdots i_{2n}}^{j_1 j_2\cdots j_{2n}}, \qquad
Z^I=Z^{i_1 i_2}\cdots Z^{i_{2n-1} i_{2n}}.
\end{equation}
First, the contractor $C_J^{\sigma_{4\mu}}$ can be written as
\begin{eqnarray}
C_J^{\sigma_{4\mu}}&=&\delta_{j_2 j_{\sigma_{4\mu}(2)}}\cdots \delta_{j_{2n} j_{\sigma_{4\mu}(2n)}} \nonumber \\ 
&=&\delta_{j_{\sigma_{4\mu}(1)} j_{\sigma_{4\mu}(2)}}\delta_{j_{\sigma_{4\mu}(3)} j_{\sigma_{4\mu}(4)}}\cdots \delta_{j_{\sigma_{4\mu}(2n-1)} j_{\sigma_{4\mu}(2n)}}\nonumber \\
&=&\delta_{k_1 k_2}\delta_{k_3 k_4}\cdots \delta_{k_{2n-1} k_{2n}} \big{(}\sigma_{4\mu}\big{)}^K_J,
\end{eqnarray}
The second line follows because $\sigma_{4\mu}(r)=r+1$ for $r$ odd. Consequently, (\ref{FOP}) becomes
\begin{eqnarray}
O_R^{\sigma_{4\mu}}&=&\frac{1}{2n!}\sum_{\beta \in S_{2n}}\chi_R(\beta) C_J^{\sigma_{4\mu}}\big{(}\beta\big{)}_I^J Z^I \nonumber \\
&=&\frac{1}{2n!}\sum_{\beta \in S_{2n}}\chi_R(\beta) \delta_{k_1 k_2}\cdots \delta_{k_{2n-1} k_{2n}} \big{(}\sigma_{4\mu}\big{)}^K_J\big{(}\beta\big{)}_I^J Z^I \nonumber \\
&=&\frac{1}{2n!}\sum_{\beta \in S_{2n}}\chi_R(\beta) \delta_{k_1 k_2}\cdots \delta_{k_{2n-1} k_{2n}} \big{(}\beta\sigma_{4\mu}\big{)}_I^K Z^I \nonumber \\
&=&\frac{1}{2n!}\sum_{\beta \in S_{2n}}\chi_R(\beta \sigma_{4\mu}^{-1}) \delta_{k_1 k_2}\cdots \delta_{k_{2n-1} k_{2n}} \big{(}\beta \big{)}_I^K Z^I. \nonumber \\
\end{eqnarray}
Since $Z$ is antisymmetric, for any $\eta\in S_n[S_2]$ we have
\begin{equation}\label{1}
 Z^{\eta(I)}=Z^{i_{\eta (1)} i_{\eta (2)}}\cdots Z^{i_{\eta (2n-1)} i_{\eta (2n)}}=\text{sgn}(\eta)Z^I, 
\end{equation}
It is now straight forward to see that
\begin{eqnarray}\label{12}
\delta_{k_1 k_2}\cdots \delta_{k_{2n-1} k_{2n}} \big{(}\xi \beta \eta \big{)}_I^K Z^I
&=& \delta_{k_1 k_2}\cdots \delta_{k_{2n-1} k_{2n}}  Z^{\eta^{-1} \beta^{-1} \xi^{-1} (K)}\nonumber \\
 &=& \delta_{k_1 k_2}\cdots \delta_{k_{2n-1} k_{2n}} Z^{\eta^{-1} \beta^{-1} (K)}\text{sgn}(\xi) \nonumber \\
 &=& \delta_{k_1 k_2}\cdots \delta_{k_{2n-1} k_{2n}} \big{(} \beta \eta \big{)}_I^K Z^I\text{sgn}(\xi) \nonumber \\
 &=& \delta_{k_1 k_2}\cdots \delta_{k_{2n-1} k_{2n}} \big{(} \beta \big{)}_I^K Z^I\text{sgn}(\xi). 
\end{eqnarray}
Consequently
\begin{eqnarray}
O_R^{\sigma_{4\mu}}&=&\frac{1}{2n!}\sum_{\beta \in S_{2n}}\chi_R(\beta \sigma_{4\mu}^{-1}) \delta_{k_1 k_2}\cdots \delta_{k_{2n-1} k_{2n}} \big{(}\beta \big{)}_I^K Z^I. \nonumber \\
&=&\frac{1}{2n!}\frac{1}{(2^n n!)^2}\sum_{\beta \in S_{2n}}\sum_{\xi, \eta \in S_n[S_2]}\chi_R(\xi \beta \eta \sigma_{4\mu}^{-1}) \delta_{k_1 k_2}\cdots \delta_{k_{2n-1} k_{2n}} \big{(}\xi \beta \eta \big{)}_I^K Z^I. \nonumber \\
&=&\frac{1}{2n!}\frac{1}{(2^n n!)^2}\sum_{\beta \in S_{2n}}\Big{[}\sum_{\xi, \eta \in S_n[S_2]}\chi_R(\xi \beta \eta \sigma_{4\mu}^{-1}) \text{sgn}(\xi)\delta_{k_1 k_2}\cdots \delta_{k_{2n-1} k_{2n}} \big{(} \beta \big{)}_I^K Z^I\Big{]}. \nonumber 
\end{eqnarray}
Now, (\ref{12}) implies that the expression within brackets defines a function on the double coset $S_n[S_2]\setminus S_{2n}/ S_n[S_2]$, 
i.e. it takes the same value for different $\sigma\in S_{2n}$ that represent the same double coset element.
Thus, we can trade the sum over $\beta\in S_{2n}$ for a sum over coset representatives which is a sum over partitions
\begin{eqnarray}
O_R^{\sigma_{4\mu}}&=&\frac{1}{2n!}\sum_{\nu \vdash n/2}\sum_{\xi, \eta \in S_n[S_2]}\frac{1}{z_{4\nu}}\chi_R(\xi \beta_{4\nu} \eta \sigma_{4\mu}^{-1}) \text{sgn}(\xi)\delta_{k_1 k_2}\cdots \delta_{k_{2n-1} k_{2n}} \big{(} \beta_{4\nu} \big{)}_I^K Z^I. \nonumber \\
&=&\frac{1}{d_R}\sum_{\nu \vdash n/2}\frac{1}{z_{4\nu}}2^{l(\mu)+l(\nu)}\chi_{R/4}(\mu)\chi_{R/4}(\nu)\delta_{k_1 k_2}\cdots \delta_{k_{2n-1} k_{2n}} \big{(} \beta_{4\nu} \big{)}_I^K Z^I,
\end{eqnarray}
To obtain the last line above we have used the mathematical identity\cite{ivanov}
\begin{equation}
\frac{d_R}{2n!}\sum_{\xi, \eta \in S_n[S_2]}\chi_R(\xi \beta_{4\nu} \eta \sigma_{4\mu}^{-1}) \text{sgn}(\xi)=2^{l(\mu)+l(\nu)}\chi_{R/4}(\mu)\chi_{R/4}(\nu).
\end{equation}
Thus, our operators become
\begin{eqnarray}
O_R^{\sigma_{4\mu}}&=&2^{l(\mu)}\frac{\chi_{R/4}(\mu)}{d_R}\sum_{\nu \vdash n/2}\frac{1}{z_{4\nu}}2^{l(\nu)}\chi_{R/4}(\nu)\delta_{k_1 k_2}\cdots \delta_{k_{2n-1} k_{2n}} \big{(} \beta_{4\nu} \big{)}_I^K Z^I \nonumber \\
&=&2^{l(\mu)}\frac{\chi_{R/4}(\mu)}{d_R}\sum_{\nu \vdash n/2}\frac{1}{z_{\nu}}2^{-l(\nu)}\chi_{R/4}(\nu)C^{\beta_{4\nu}}_I Z^I \nonumber \\
&=&2^{l(\mu)}\frac{\chi_{R/4}(\mu)}{d_R}\frac{1}{(n/2)!}\sum_{\sigma \in S_{n/2}}2^{-l(\sigma)}\chi_{R/4}(\sigma)\text{Tr}\big{(}\sigma (Z^2)^{\otimes \frac{n}{2}}\big{)}. 
\end{eqnarray}
For the special case $\mu=(1^{n/2})$ we find
\begin{equation}\label{def3}
O_R=2^{n/2}\frac{d_{R/4}}{d_R}\frac{1}{(n/2)!}\sum_{\sigma \in S_{n/2}}2^{-l(\sigma)}\chi_{R/4}(\sigma)\text{Tr}\big{(}\sigma (Z^2)^{\otimes \frac{n}{2}}\big{)}.
\end{equation}
which completes the demonstration.

\section{Correlation Functions}

To test the result (\ref{completeresult}) we have studied a number of correlation functions using Mathematica and analytic techniques.
Consider first correlators of the form $\langle {\rm Tr}(Z^2)^p {\rm Tr}(\bar{Z}^2)^p\rangle$.
Introduce the notation
\bea
   A_p=\langle {\rm Tr}(Z^2)^p {\rm Tr}(\bar{Z}^2)^p\rangle
\eea
By studying Wick contractions (or equivalently Schwinger-Dyson equations) it is not hard to obtain the following
recursion relation
\bea
  A_p = (16p(p-1)+4pN(N-1))A_{p-1}
\eea
Since $A_0=1$ we can easily generate explicit expression for $A_p$. For example
\bea
  A_5&=&122880N^{10}-614400N^9+6144000N^8-20889600N^7+98918400N^6-226222080N^5\cr
      && +604569600N^4-855244800N^3+1148190720N^2-754974720N
\eea
Introduce the short hand
\bea
   \prod_{i\in {\rm odd\, boxes\, in\,}R} c_i  =   f^{\rm odd}_R
\eea
Our formula (\ref{completeresult}) says
\bea
A_5=2^{10}\left(
f^{\rm odd}_{\tiny \yng(10,10)}
+16 f^{\rm odd}_{\tiny \yng(8,8,2,2)}
+25 f^{\rm odd}_{\tiny \yng(6,6,4,4)}
+36 f^{\rm odd}_{\tiny \yng(6,6,2,2,2,2)}
+25f^{\rm odd}_{\tiny \yng(4,4,4,4,2,2)}
+16f^{\rm odd}_{\tiny \yng(4,4,2,2,2,2,2,2)}
+f^{\rm odd}_{\tiny \yng(2,2,2,2,2,2,2,2,2,2)}\right)
\eea
which is indeed correct.

Next, consider correlators of the form $\langle {\rm Tr}(X^2)^p{\rm Tr}(X^4){\rm Tr}(\bar{X}^2)^{p+2}\rangle$.
Introduce the notation
\bea
   B_p=\langle {\rm Tr}(X^2)^p{\rm Tr}(X^4){\rm Tr}(\bar{X}^2)^{p+2}\rangle
\eea
Again by studying Wick contractions (or equivalently Schwinger-Dyson equations) it is not hard
to obtain the following recursion relation
\bea
  B_p=16(2p+4)(2p+2)(2p)(2p-2)B_{p-2}+2(2p+4)(N-1)A_{p+1}\cr
      +[4N^2(N-1)(2p+4)(2p+2)+(2p+4)(2p+2)(2p)8(3N-1)]A_p
\eea
We now easily find, for example,
\bea
  B_3 = 245760N^9-1105920N^8+10813440N^7 -32686080N^6+142786560N^5\cr
       -277831680N^4+586383360N^3-617349120N^2+188743680N
\eea
Our formula (\ref{completeresult}) says
\bea
B_3=2^9\left(
f^{\rm odd}_{\tiny \yng(10,10)}
+8 f^{\rm odd}_{\tiny \yng(8,8,2,2)}
+5 f^{\rm odd}_{\tiny \yng(6,6,4,4)}
+0
-5f^{\rm odd}_{\tiny \yng(4,4,4,4,2,2)}
-8f^{\rm odd}_{\tiny \yng(4,4,2,2,2,2,2,2)}
-f^{\rm odd}_{\tiny \yng(2,2,2,2,2,2,2,2,2,2)}\right)
\eea
which is indeed correct.

Finally, we have also performed a complete check in Mathematica for the full set of operators that can be built using
8 fields. The Mathematica and analytic results from (\ref{completeresult}) are again in complete agreement.
A study of the $Sp(N)$ coorelators has also been performed to confirm (\ref{spncorrelators}) using Mathematica.

\section{Jacobians}\label{jacs}

For single matrix models, since we are interested in the dynamics of gauge invariant observables, we can employ an eigenvalue description. 
The idea is to write the path integral as an integral over the eigenvalues and some angles. 
We then integrate out the angles.
The resulting measure is nontrivial.
With a slight abuse of language, we refer to the measure as a Jacobian.
In this Appendix we would like to compute the Jacobian for a single complex $SO(N)$ matrix and for a single complex $Sp(N)$ matrix.
To the best of our knowledge, these are new results.

The approach we employ for determining the measure, is first to compute it for a real matrix and then use this result to
guess the answer for the complex matrix.
We check this guess by verifying that we get the correct answer for any correlation function we compute.
To determine the Jacobian for a real (or Hermitian) matrix, we require that the Schwinger-Dyson equations in the original (matrix) 
variables agree with the Schwinger-Dyson equations in the eigenvalue variables. 
This implies a differential equation for the Jacobian which we solve.
Although the Jacobians we compute in this way are all known, we have given our derivation since it
seems to be new and is simpler than existing derivations. 
To start, we illustrate the method with the $U(N)$ matrix model and then move on to $SO(N)$ and $Sp(N)$.

The use of Schwinger-Dyson equations to determine a Jacobian in this way was pioneered in collective field theory\cite{Jevicki:1980zg}.
See \cite{Jevicki:1993rr} for applications to multi matrix models and \cite{deMelloKoch:1996mj} for applications to vector models.

\subsection{$U(N)$ Matrix Models}

The main goal of this subsection is to illustrate how we compute the Jacobian using Schwinger-Dyson equations and also
to illustrate the close connection between the Jacobian for the real (or Hermitian) matrix and the complex version.

Consider the matrix model for a single matrix $X$ living in the Lie algebra u$(N)$, i.e. $X$ is Hermitian. 
Rewrite the Schwinger-Dyson equation
\bea
0&=&\int [dX]{d\over dX_{ij}}\left( [X^{n-1}]_{ij}e^{-{1\over 2}{\rm Tr} (X^2)}\right)\cr
 &=&\int [dX] \left( \sum_{r=0}^{n-2}{\rm Tr}[X^{n-r-2}] {\rm Tr}[X^{r}] -{\rm Tr}[X^n]\right) e^{-{1\over 2}{\rm Tr} (X^2)}
\eea
in terms of eigenvalue variables to obtain
\bea
 0 = \int [d\lambda] J(\lambda)\left(\sum_{r=0}^{n-2}\sum_{i,j=1}^N \lambda_i^{n-r-2} \lambda_j^r
         -\sum_{i=1}^N\lambda_i^n\right) e^{-{1\over 2}\sum_{l=1}^N\lambda_l^2}
\eea
where $J(\lambda)$ is the Jacobian we want to determine.
After performing the sum over $r$ we have
\bea
0=\int [d\lambda] J(\lambda)\left( 2 \sum_{i,j=1\, i\ne j}^N {\lambda_i^{n-1}\over \lambda_i-\lambda_j}
      + (n-1)\sum_{i=1}^N\lambda_i^{n-2}-\sum_{i=1}^N\lambda_i^n\right) e^{-{1\over 2}\sum_{l=1}^N\lambda_l^2}
\label{firstSD}
\eea
Now, work directly in the eigenvalue variables
\bea
0&=&\int[d\lambda] \sum_{i=1}^N{\partial\over\partial\lambda_i}\left( \lambda_i^{n-1}J(\lambda )e^{-{1\over 2}\sum_{l=1}^N\lambda_l^2}\right)\cr
&=&\int[d\lambda] J(\lambda)\left(\sum_{i=1}^N \lambda_i^{n-1}{\partial \log J(\lambda)\over\partial\lambda_i}
+(n-1)\sum_{i=1}^{N}\lambda_i^{n-2}-\sum_{i=1}^N\lambda_i^n\right) e^{-{1\over 2}\sum_{l=1}^N\lambda_l^2}
\label{secondSD}
\eea
Comparing (\ref{firstSD}) and the second line of (\ref{secondSD}) we learn that
\bea 
 {\partial \log J(\lambda)\over\partial\lambda_i}=2 \sum_{j=1\, i\ne j}^N
        {1\over \lambda_i-\lambda_j}
    \label{jdiff}
\eea
This clearly implies that
\bea
   J(\lambda )=\prod_{i<j} (\lambda_i-\lambda_j)^2\equiv (\Delta (\lambda))^2
   \label{realJac}
\eea
which is indeed the correct result.

The argument given above is not quite rigorous. Indeed, to really prove that we will get the correct answer for {\it any} gauge
invariant observable in the theory, we should have considered the Schwinger-Dyson equation which follows from
\bea
   0=\int [dX]{d\over dX_{ij}}\left( [X^{n-1}]_{ij} {\cal F}e^{-S}\right)\cr
\eea
with $S={1\over 2}{\rm Tr} (X^2)$ and with ${\cal F}$ any gauge invariant observable in the theory. 
This gives the complete set of Schwinger-Dyson equations of the theory.
This much more general argument continues to imply (\ref{jdiff}). 
Further, we could also have included interactions (i.e. considered ``actions'' $S$ with higher than quadratic terms);
one again finds (\ref{jdiff}). 
In what follows, for the sake of simplicity, we will continue to work in the free theory and simply set ${\cal F}$ to one.
The reader should bear in mind that adding interactions and/or considering the most general ${\cal F}$ do not change our
results.

We will need a generalization of the above result. 
Consider a matrix model of two Hermitian matrices $X$ and $Y$.
Collect these into the complex matrix $Z={1\over\sqrt{2}}(X+iY)$ and its Hermitian conjugate $Z^\dagger$.
We are interested in correlation functions of traces of $Z$s or traces of $Z^\dagger$s, but not of traces 
in which both $Z$ and $Z^\dagger$ appear in the same trace. 
Denote the eigenvalues of $Z$ by $z_i$.
The eigenvalues of $Z^\dagger$ are then $\bar{z}_i$.
The Jacobian in this case is\cite{Ginibre:1965zz}
\bea
   J=\Delta (z)\Delta (\bar{z})
   \label{cmplxJac}
\eea
The relation between (\ref{realJac}) and (\ref{cmplxJac}) is striking; we will find a similar relation between the 
Jacobians for the real and complex $SO(N)$ and $Sp(N)$ theories below. 

\subsection{$SO(N)$ Matrix Models}

We will now consider the matrix model relevant for $SO(N)$ gauge theory. We consider the case that $N$ is even and start with
a single real $N\times N$ antisymmetric matrix $X$. In this case, we can, with a unitary transformation, bring $X$ into a
block diagonal form
\bea
X=\left[
\begin{array}{ccccc}
0        &x_1     &0     &0     &        \\
-x_1     &0       &0     &0     &\cdots  \\
0        &0       &0     &x_2   &        \\
0        &0       &-x_2  &0     &        \\
         &\vdots  &      &      &
\end{array}
\right]
\eea
Using this explicit form for the matrix $X$ we easily find ${\rm Tr}(X^{2J+1})=0$ and
\bea
  {\rm Tr}(X^{2J})= 2(-1)^J\sum_{i=1}^{N\over 2} x_i^{2J}
  \label{traceofX}
\eea
Since $X$ is an antisymmetric matrix, elements above and below the diagonal are related by a sign. Consequently
\bea
   {d X_{kl}\over dX_{ij}}=\delta_{ik}\delta_{jl}-\delta_{il}\delta_{jk}
\eea
To get some practice with this derivative and since we need it in what follows, consider
\bea
  {d\over dX_{ij}} (X^{2J-1})_{ij}&=&\sum_{r=0}^{2J-2} (X^r)_{ik} {d X_{kl}\over dX_{ij}} (X^{2J-r-2})_{lj}\cr
                                  &=&\sum_{r=0}^{2J-2}(X^r)_{ik}(\delta_{ik}\delta_{jl}-\delta_{il}\delta_{kj})(X^{2J-r-2})_{lj}\cr
                                  &=&\sum_{r=0}^{2J-2}\left[ {\rm Tr}(X^r){\rm Tr}(X^{2J-r-2})-{\rm Tr}((X^T)^r X^{2J-r-2})\right]\cr
                                  &=&\sum_{r=0}^{J-1}{\rm Tr}(X^{2r}){\rm Tr}(X^{2J-2r-2})-\sum_{r=0}^{2J-2}(-1)^r {\rm Tr}(X^{2J-2})\cr
                                  &=&\sum_{r=0}^{J-1}{\rm Tr}(X^{2r}){\rm Tr}(X^{2J-2r-2})-{\rm Tr}(X^{2J-2})
\eea
Now, consider the Schwinger-Dyson equation
\bea
   0=\int [dX]{d\over dX_{ij}}\left( (X^{2J-1})_{ij}e^{{1\over 2}{\rm Tr}(X^2)}\right)\cr
\eea
which implies
\bea
  \langle \sum_{r=0}^{J-1}{\rm Tr}(X^{2r}){\rm Tr}(X^{2J-2r-2})-{\rm Tr}(X^{2J-2})\rangle = - 2\langle {\rm Tr} (X^{2J})\rangle
\eea
Writing this in terms of eigenvalues gives
\bea
    \langle 4 (-1)^{J+1} \sum_{r=0}^{J-1}\sum_{i,j=1}^{N\over 2} x_i^{2r}x_j^{2J-2r-2}-2(-1)^{J+1}\sum_{i=1}^{N\over 2} x_i^{2J-2}\rangle
   = -4(-1)^J\langle \sum_{i=1}^{N\over 2} x_i^{2J}\rangle
\eea
or, after summing over $r$
\bea
     \langle 8 \sum_{i,j=1, i\ne j}^{N\over 2} {x_i^{2J}\over x_i^2 -x_j^2}+2(2J-1)\sum_{i=1}^{N\over 2} x_i^{2J-2}\rangle
     = 4\langle \sum_{i=1}^{N\over 2} x_i^{2J}\rangle
     \label{firstson}
\eea
Now, working in terms of the eigenvalue variables we have
\bea
  0=\int [dx]\sum_{i=1}^{N\over 2} {d\over dx_i}\left( J x_i^{2J-1}e^{-\sum_{j=1}^{N\over 2} x_j^2}\right)
\eea
which becomes
\bea
  0=\int [dx] J e^{-\sum_{j=1}^{N\over 2} x_j^2} \sum_{i=1}^{N\over 2}\left( x_i^{2J-1}{d\log J\over dx_i}+(2J-1)x_i^{2J-2}-2x_i^{2J}\right)
  \label{secondson}
\eea
Comparing (\ref{firstson}) and (\ref{secondson}) we have
\bea
    {d\log J\over dx_i}=\sum_{i,j=1,j\ne i}^{N\over 2} {4x_i\over x_i^2-x_j^2}
\eea
which is solved by
\bea
   J(x)=\prod_{i<j=1}^{N\over 2} (x_i^2 -x_j^2)^2=(\Delta(x^2))^2
\eea
This is the correct answer. 

We now consider a model of two real $SO(N)$ matrices $X$ and $Y$. Introduce the complex matrix $Z={1\over\sqrt{2}}(X+iY)$ and its Hermitian
conjugate $Z^\dagger$. Denote the eigenvalues of $Z$ by $z_i$. It follows that the eigenvalues of $Z^\dagger$ are given by
complex conjugation as $\bar{z}_i$. We are interested in computing correlation functions of multi trace operators that have only $Z$s or
$Z^\dagger$s in each trace. Based on the relation between (\ref{realJac}) and (\ref{cmplxJac}), we conjecture that the Jacobian
is given by
\bea
  J(z,\bar{z})=\Delta (z^2)\Delta (\bar{z}^2)
  \label{cmplxson}
\eea
To prove that this guess is correct, we will compute two point functions of our operators $\chi_R^0(Z)$ and show that they
agree perfectly with our result (\ref{twopointsfornew}) computed directly in the matrix model. Since the most general
multi trace operators that have only $Z$s or $Z^\dagger$s appearing in each trace can be written as a linear combination
of the $\chi_R^0(Z)$ and $\chi_R^0(Z^\dagger)$, this proves that (\ref{cmplxson}) is the correct Jacobian.

Correlation functions in the eigenvalue basis are computed using
\bea
  \langle \cdots\rangle = {\cal N}\int [dzd\bar{z}] J(z,\bar{z})e^{-\sum_{i=1}^{N\over 2}z_i\bar{z}_i}\,\,\cdots
\eea
We will fix the normalization by requiring that $\langle 1\rangle =1$. 
To perform this computation, note that we can write
\bea
  \Delta (z^2)&=&\prod_{i<j=1}^{N\over 2} (z_i^2 -z_j^2)\cr
              &=&{\rm det}\left[
\begin{array}{cccc}
z_1^{N-2}          &z_2^{N-2} &\cdots  &z_{N\over 2}^{N-2}\cr
z_1^{N-4}          &z_2^{N-4} &\cdots  &z_{N\over 2}^{N-4}\cr
\vdots             &\vdots    &\vdots  &\vdots            \cr
z_1^2              &z_2^2     &\cdots  &z_{N\over 2}^2    \cr
1                  &1         &\cdots  &1
\end{array}
\right]\cr
&=&\epsilon^{i_1 i_2 \cdots i_{N\over 2}}z_1^{N-2i_1} z_2^{N-2i_2}\cdots z_{N\over 2}^{N-2i_{N\over 2}}
\eea
We will make frequent use of the identity
\bea
   \int dzd\bar{z} \,\, z^n\bar{z}^m e^{-z\bar{z}}\,\,=\,\,\pi\delta^{mn}n!
\eea
It is now straight forward to verify that
\bea
  {\cal N}={1\over \pi^{N\over 2}\left({N\over 2}\right)!\prod_{i=1}^{N\over 2} (N-2i)!}
\eea
Next, using the identity
\bea
   \Delta (z^2)\chi_R^0(Z)=\epsilon^{i_1 i_2 \cdots i_{N\over 2}}z_1^{r_{i_1}+N-2i_1} z_2^{r_{i_2}+N-2i_2}\cdots 
                                                                 z_{N\over 2}^{r_{i_{N\over 2}}+N-2i_{N\over 2}}
\eea
we easily find
\bea
  \langle \chi_R^0(Z)\chi_S^0(Z^\dagger )\rangle =
{\prod_{i=1}^{N\over 2} (r_i+N-2i)!
\over
\prod_{i=1}^{N\over 2} (N-2i)!}
=\delta_{RS}f_{R,{\rm odd}}
\eea
in perfect agreement with (\ref{twopointsfornew}). A simple generalization of this argument shows that we correctly
reproduce the correlators for odd $N$ too. For the odd $N$ case the Jacobian is
\bea
  J(z,\bar{z})=\tilde{\Delta} (z)\tilde{\Delta} (\bar{z})
\eea
where
\bea
\tilde{\Delta} (z)=\prod_{k=1}^{N-1\over 2}z_k\prod_{1\le i<j\le {N-1\over 2}} (z_i^2-z_j^2)
\eea

\subsection{$Sp(N)$ Matrix Models}

Finally, consider the matrix model relevant for $Sp(N)$ gauge theory. In this case, $N$ is even. Start with
a single real $N\times N$ matrix $X$ in the Lie algebra of $Sp(N)$ which hence obeys
\bea
  X^T=\Omega X\Omega
  \label{notindependent}
\eea
It is easy to see that
\bea
  0=\det (\lambda{\bf 1}-X)=\det (\lambda{\bf 1}-X^T)=\det (\lambda{\bf 1}-\Omega X\Omega)=\det (\lambda{\bf 1}+X)
\eea
so that the eigenvalues of $X$ come in pairs $\pm x_i$, $i=1,2,...,{N\over 2}$.
Consequently, we find ${\rm Tr}(X^{2J+1})=0$ and
\bea
  {\rm Tr}(X^{2J})= 2\sum_{i=1}^{N\over 2} x_i^{2J}
  \label{traceofspX}
\eea
The relation (\ref{notindependent}) implies
\bea
   {d X_{kl}\over dX_{ij}}=\delta_{ik}\delta_{jl}+\Omega_{il}\Omega_{kj}
\eea
It is now straight forward to verify that
\bea
  {d\over dX_{ij}} (X^{2J-1})_{ij}&=&\sum_{r=0}^{J-1}{\rm Tr}(X^{2r}){\rm Tr}(X^{2J-2r-2})+{\rm Tr}(X^{2J-2})
\eea
Now, consider the Schwinger-Dyson equation
\bea
   0=\int [dX]{d\over dX_{ij}}\left( (X^{2J-1})_{ij}e^{-{1\over 2}{\rm Tr}(X^2)}\right)\cr
\eea
which implies
\bea
  \langle \sum_{r=0}^{J-1}{\rm Tr}(X^{2r}){\rm Tr}(X^{2J-2r-2})+{\rm Tr}(X^{2J-2})\rangle = 2\langle {\rm Tr} (X^{2J})\rangle
\eea
Writing this in terms of eigenvalues and summing over $r$ gives
\bea
     \langle 8 \sum_{i,j=1, i\ne j}^{N\over 2} {x_i^{2J}\over x_i^2 -x_j^2}+2(2J+1)\sum_{i=1}^{N\over 2} x_i^{2J-2}\rangle
     = 4\langle \sum_{i=1}^{N\over 2} x_i^{2J}\rangle
     \label{firstson}
\eea
Now, working in terms of the eigenvalue variables we have
\bea
  0=\int [dx]\sum_{i=1}^{N\over 2} {d\over dx_i}\left( J x_i^{2J-1}e^{-\sum_{j=1}^{N\over 2} x_j^2}\right)
\eea
which becomes
\bea
  0=\int [dx] J e^{-\sum_{j=1}^{N\over 2} x_j^2} \sum_{i=1}^{N\over 2}\left( x_i^{2J-1}{d\log J\over dx_i}+(2J-1)x_i^{2J-2}-2x_i^{2J}\right)
  \label{secondson}
\eea
Comparing (\ref{firstson}) and (\ref{secondson}) we have
\bea
    {d\log J\over dx_i}=\sum_{i,j=1,j\ne i}^{N\over 2} {4 x_i\over x_i^2-x_j^2}+{2\over x_i}
\eea
which is solved by
\bea
   J(x_i) =\left(\prod_{k=1}^{N\over 2} x_k \prod_{i<j=1}^{N\over 2} (x_i^2 -x_j^2)\right)^2\equiv \tilde{\Delta}(x)^2
\eea
This is the correct answer. 

We now consider a model of two real $Sp(N)$ matrices $X$ and $Y$.
Introduce the complex matrix $Z={1\over\sqrt{2}}(X+iY)$ and its Hermitian conjugate $Z^\dagger$. 
Denote the eigenvalues of $Z$ by $z_i$ and of $Z^\dagger$ by $\bar{z}_i$. 
Again we are interested in computing correlation functions of multitrace operators that have only $Z$s or $Z^\dagger$s in each trace. 
Given the relation between (\ref{realJac}) and (\ref{cmplxJac}), we conjecture that the Jacobian is given by
\bea
  J(z,\bar{z})=\tilde{\Delta} (z^2)\tilde{\Delta} (\bar{z}^2)
  \label{cmplxspn}
\eea
To prove that this guess is correct, we will again compute two point functions of our operators $\chi_R^0(Z)$ and show that they
agree perfectly with our result (\ref{sptwopoint}) computed directly in the matrix model. 

Correlation functions in the eigenvalue basis are computed using
\bea
  \langle \cdots\rangle = {\cal N}\int [dzd\bar{z}] J(z,\bar{z})e^{-\sum_{i=1}^{N\over 2}z_i\bar{z}_i}\cdots
\eea
Again fix the normalization by requiring that $\langle 1\rangle =1$. 
Note that
\bea
  \tilde{\Delta} (z^2)&=&\prod_{i<j=1}^{N\over 2} (z_i^2 -z_j^2)\prod_{k=1}^{N\over 2}z_k\cr              
  &=&\epsilon^{i_1 i_2 \cdots i_{N\over 2}}z_1^{N-2i_1+1} z_2^{N-2i_2+1}\cdots z_{N\over 2}^{N-2i_{N\over 2}+1}
\eea
It is now straight forward to verify that
\bea
  {\cal N}={1\over \pi^{N\over 2}\left({N\over 2}\right)!\prod_{i=1}^{N\over 2} (N-2i+1)!}
\eea
Next, using the identity
\bea
   \Delta (z^2)\chi_R^0(Z)=\epsilon^{i_1 i_2 \cdots i_{N\over 2}}z_1^{r_{i_1}+N-2i_1+1} z_2^{r_{i_2}+N-2i_2+1}\cdots 
                                                                 z_{N\over 2}^{r_{i_{N\over 2}}+N-2i_{N\over 2}+1}
\eea
we easily find
\bea
  \langle \chi_R^0(Z)\chi_S^0(Z^\dagger )\rangle =
{\prod_{i=1}^{N\over 2} (r_i+N-2i+1)!
\over
\prod_{i=1}^{N\over 2} (N-2i+1)!}
=\delta_{RS}f_{R,{\rm even}}
\eea
in perfect agreement with (\ref{sptwopoint}). 

\end{appendix}

\end{document}